\documentclass[a4paper,11pt]{article}
\usepackage{jheppub} 
\usepackage{xcolor}
\usepackage{nicefrac} %
\usepackage[stable]{footmisc}
\usepackage[font=small,labelfont=bf,tableposition=top]{caption}
\DeclareCaptionLabelFormat{andtable}{#1~#2  \&  \tablename~\thetable}
\usepackage{soul}
\usepackage{threeparttable}
\usepackage[subpreambles=true]{standalone}
\usepackage{booktabs}
\usepackage{diagbox}
\renewcommand{\emph}[1]{#1}

\arxivnumber{}
\title{Effective strings and particles interacting in 3D: \\the Ising model}

\author[a,b,c]{J. M. Viana Parente Lopes}
\author[a,b,c,d]{José Matos}
\author[d]{Joao Penedones}
\affiliation[a]{Associate Laboratory LaPMET, 4169-007 Porto, Portugal.}
\affiliation[b]{Departamento de Física e Astronomia, Faculdade de Ciências, Universidade do Porto, rua do Campo Alegre s/n, 4169-007 Porto, Portugal.}
\affiliation[c]{Centro de Física das Universidades do Minho e Porto (CF-UM-PT), Departamento de Física e Astronomia, Faculdade de Ciências, Universidade do Porto, 4169-007 Porto, Portugal.}
\affiliation[d]{Fields and Strings Laboratory, Institute of Physics, École Polytechnique Fédérale de Lausanne (EPFL), Route de la Sorge, CH-1015 Lausanne, Switzerland}

\emailAdd{jose.bouradematos@gmail.com}

\author{}

\abstract{
 We study how a fluctuating domain wall in three dimensions modifies bulk observables in a gapped phase. We introduce an effective interaction between the wall and the lightest bulk massive mode, and identify the regime in which this description is controlled: nearly on-shell bulk exchange with small momentum along the wall. In this regime, several  observables are controlled by a renormalized dimensionless coupling $\lambda$, including the large-$L_z$ correction to the wall free energy and the large-$|x_\perp|$ tail of two-point functions in the presence of the wall. Other observables, such as one-point functions and two-point functions in the nearby-regime, retain non-universal dependence on operator data and on the bulk spectral density. We test the universal kinematic consequences of wall fluctuations, and find good agreement with the predicted rough-wall broadening and nearby Gaussian behavior in Monte Carlo simulations of the 3D Ising model with anti-periodic boundary conditions. 
}

\begin{document}
\maketitle

\section{Introduction}

Quantum field theories can support extended excitations in addition to ordinary particles.
In confining pure gauge theories, there are particles (glueballs)  and strings (flux-tubes).\footnote{There are also string junctions \cite{Komargodski:2024swh, Caselle:2025elf}.}  
In the ferromagnetic phase of the 3D Ising field theory, there are particles and strings (domain walls).
At long distances, the dynamics of such strings is captured by Effective String Theory (EST)
\cite{Nambu:1974zg,Luscher:1980ac,Polchinski:1991ax,Caselle:1992ue,Caselle:1994zt,Billo:2006zg,Caselle:2007yc,Dubovsky:2012sh,Aharony:2013ipa,Dubovsky:2013gi,Dubovsky:2015zey,Dubovsky:2014fma,Conkey:2016qju,Beratto:2019bap}.
In this work, we initiate the study of the interactions between  particles and strings.

We will only need the leading term of the EST in 3d, namely the Gaussian theory of the single transverse Goldstone mode. In the static gauge $X^\mu(x )=(x^1 ,x^2 ,\pi(x ))$, the worldsheet action in the low momentum limit %
reduces to
\begin{equation}
S_{\rm EST}= \frac{\sigma}{2}\int d^2x \,(\partial\pi)^2  \;+\dots,
\label{eq:EST-free}
\end{equation}
where $\sigma$ is the string tension.
We refer to $\pi$ as the \emph{branon}. 
In addition, we consider a massive bulk scalar operator, $\phi$, that can create the lightest bulk particle from the vacuum.
We model the interaction between the string and the particle by the simplest diffeomorphism and target space Poincaré-invariant action:
\begin{equation}
S_{\textrm{int}}= \lambda_0 \int d^{2}x  \sqrt{h}\,\phi\left(x ,\pi(x)\right)\,,\label{eq:interaction}
\end{equation}
where $h_{\mu\nu}=\delta_{\mu\nu} +\partial_\mu\pi \partial_\nu \pi $ is the induced metric and we introduced the bare coupling $\lambda_0$.
At this point, the knowledgeable reader will be surprised because she knows that, generically, the UV cutoff of the EST is of the order of the mass of the bulk particle $m\sim \sqrt{\sigma}$.
We will argue that this setup makes sense for observables that are dominated by the exchange of nearly on-shell bulk particles with low momentum along the worldsheet.

In sec.~\ref{sec:theory}, we develop 
this framework and make predictions for several observables: finite transverse volume corrections to the wall free energy,
glueball scattering off the wall,
and bulk one-point and two-point functions in the presence of the wall.
In sec.~\ref{sec:theory-geom}, we use a simple geometric description of the fluctuating domain wall to make similar predictions for one-point and two-point functions of bulk local operators.
In sec.~\ref{sec:numeric} we test some of the previous predictions in the 3D Ising model using Monte Carlo simulations.
Finally, we conclude in sec. \ref{sec:discussion}
with comments about applications to flux tubes in gauge theories and other future directions.

\section{Effective field theory for particle-wall interactions}\label{sec:theory}

When the bulk theory is gapped, coupling a massive bulk excitation to a fluctuating string is not, in general, a problem that can be treated within the \emph{infrared} worldsheet EFT. 
Generically, the lightest bulk mass is at the string scale, $m^{2}\sim\sigma$ (see e.g.\ Table~34 of \cite{Athenodorou:2021qvs}).\footnote{For an example with a
parametrically lighter bulk mode relative to $\sqrt{\sigma}$, see \cite{Aharony:2024ctf}.} 
In this situation, 
 real-time particle-string processes probe energies at the  cutoff scale of the derivative expansion. A concrete way to see this is to imagine a massive bulk particle impinging perpendicularly on the string and being absorbed into two branons. Energy conservation then forces each branon to carry energy $> m/2\sim \sqrt{\sigma}$, which is the cutoff scale of the EST. 

The observables of interest to us are the ones that  probe \emph{long wavelengths along the string}. In Euclidean signature, this corresponds to momenta
\begin{equation}
p^\mu=(q,\pm i\sqrt{m^2+q^2})\,,
\qquad q\in \mathbb{R}^2\,,
\qquad q^2 \ll m^2\,,
\label{eq:glueballmomentum}
\end{equation}
so that the momentum parallel to the wall is small even though  $p^2=-m^2$ is on-shell. As we shall see, momenta of this type appear, for example, in finite transverse volume corrections to the string tension.
In this regime,  the leading interaction is given by \eqref{eq:interaction}.

We normalize $\phi$ so that 
\begin{align}
    \langle \phi(x) \phi(y)\rangle_{\mathbb{R}^3} = \int_0^\infty ds \rho(s)  \int \frac{d^3k}{(2\pi)^3} \frac{e^{ik\cdot(x-y)}}{k^2+s}\,,\qquad
    \rho(s) = \delta(s-m^2) + \tilde{\rho}(s) \,,
\end{align}
where the support of $\tilde{\rho}$ is for $s\ge \tilde{m}^2>m^2$.
The long distance behavior is
\begin{align}
    \langle \phi(x) \phi(0)\rangle_{\mathbb{R}^3} \approx  \int \frac{d^3k}{(2\pi)^3} \frac{e^{ik\cdot x}}{k^2+m^2} = \frac{e^{-m |x|}}{4\pi |x|}\equiv G(x) \,.
\end{align}
In this section, we will study several observables that are independent of the spectral density $\rho(s)$.

\begin{figure}[t]
    \centering
    \includegraphics[width=0.49\linewidth]{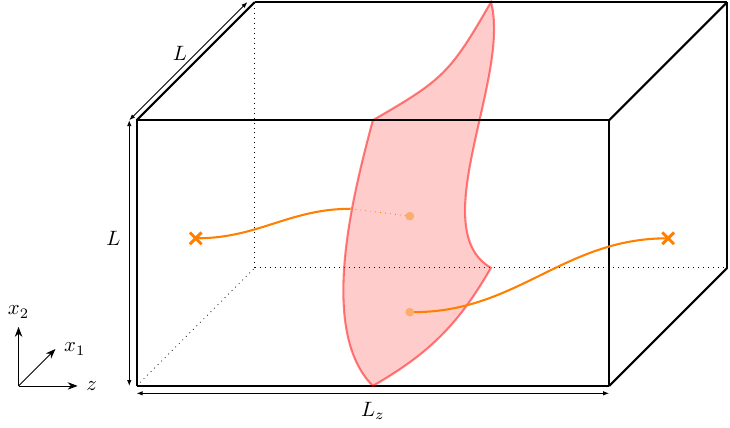}    \includegraphics[width=0.49\linewidth]{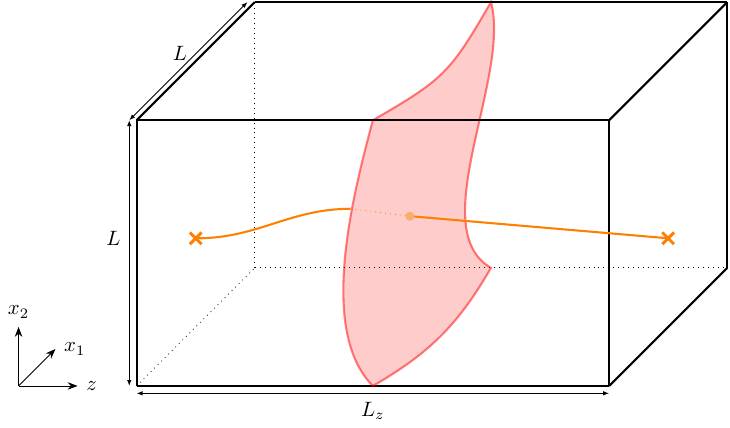}
    \caption{Left) diagrams contributing at $\mathcal{O}(\lambda^2)$ to the finite-$L_z$ correction of the wall free energy for the linear coupling in eq.~\eqref{eq:interaction}; right) diagrams contributing at $\mathcal{O}(\lambda_2)$ to the finite-$L_z$ correction of the wall free energy for the quadratic coupling in eq.~\eqref{eq:quadratic interaction}. The crosses show how the propagators are identified due to the periodic boundary conditions along the $z$ direction. }
    \label{fig:cubic diagrams}
\end{figure}

\subsection{Finite transverse volume corrections to the string tension}\label{sec: finite transverse volume corrections}

We consider a fluctuating toroidal wall of large area $A$ with a periodic transverse direction of length $L_z$ (Fig.~\ref{fig:cubic diagrams}). We define the effective string tension as the free energy per unit area,
\begin{equation}
    \sigma_{\text{eff}}(L_z)\equiv \lim_{A\to\infty} \dfrac{F(A,L_z)}{A}.
\end{equation}
Pure EST does not generate transverse finite-size dependence at $A\to\infty$   \cite{Lima:2025sqa},
\begin{equation}
\sigma_{\text{eff}}^{\text{EST}}(L_z)\equiv
\lim_{A\to\infty}\frac{F^{\text{EST}}( A,L_z)}{A}=\sigma
\qquad
(\text{with $L_z$ fixed}).
\end{equation}
Therefore, the $L_z$ dependence of $\sigma_{\text{eff}}$  is due only to interactions with bulk particles.

The leading contribution arises at second order in $S_{\text{int}}$, and can be written as 
\begin{equation}
\Delta F \;\approx\;-
\frac{1}{2} \lambda_0^2 \int d^2x \, d^2y \,
\Big[
\big\langle \phi(x ,\pi(x ))\,\phi(y ,\pi(y ))\big\rangle_{L_z}
-
\big\langle \phi(x ,\pi(x ))\,\phi(y ,\pi(y ))\big\rangle_{\infty}
\Big],
\label{eq:leading free energy finite lz corrections full}
\end{equation}
where the subscripts $L_z$ and $\infty$ denote bulk propagation with finite and infinite transverse extent, respectively. The subtraction removes  the infinite-$L_z$ free energy. We assume  $(\tilde{m}-m)L_z \gg 1$, so that we can approximate the full propagator by the free propagator.

The finite-$L_z$ free propagator admits the standard image charge expansion,
\begin{equation}
G_{L_z}(x_\parallel,x_\perp)=\sum_{n\in\mathbb{Z}}G\!\left(x_\parallel,\,x_\perp+nL_z\right).
\end{equation}
Because the bulk mode is massive, $G$ decays exponentially at large separations and the correction at large $L_z$ is dominated by the first windings $n=\pm1$ (for general finite-volume corrections in gapped theories see \cite{luscher:1985dn}).
Therefore,
\begin{align}
\Delta F & \approx -\lambda_0^2 \int d^{2}xd^{2}y\left\langle G\left(x-y,\pi\left(x\right)-\pi\left(y\right)+L_{z}\right)\right\rangle _{\pi}\nonumber \\
 & =-\lambda_0^2\int d^{2}xd^{2}ydwG\left(x-y,w+L_{z}\right)\left\langle \delta\left(w-\pi\left(x\right)+\pi\left(y\right)\right)\right\rangle _{\pi}.\label{eq:free energy in finite volume step 1}
\end{align}

The expectation value of the Dirac delta is evaluated by decomposing it in Fourier space, where it can straightforwardly be averaged over the branon field
\begin{align}
\left\langle \delta\left(w-\pi\left(x\right)+\pi\left(y\right)\right)\right\rangle _{\pi} & =
\int\dfrac{dk}{2\pi}e^{ikw}\langle 
e^{-ik\pi(x)}e^{ik\pi(y)}\rangle =\int\dfrac{dk}{2\pi}e^{ikw}e^{-\frac{k^{2}}{2}d^2\left(x-y\right) }\nonumber\\
 & =\dfrac{1}{\sqrt{2\pi }d\left(x-y\right)}\exp
 \left[-\dfrac{w^{2}}{2d^2\left(x-y\right)}\right].
 \label{eq:DiracdeltaGaussian}
\end{align}
The general picture is clear: averaging over the branon is equivalent to a gaussian smearing in the relative displacement.
Here $d^2$ is the variance of the domain wall transverse fluctuations
\begin{equation}
d^{2}(x )\;\equiv\;\big\langle\big[\pi(x)-\pi(0)\big]^{2}\big\rangle\,.
\end{equation}
For a free branon on a large torus, one finds \cite{Polchinski:1998rq}
\begin{equation}
d^{2}(x)\approx \frac{1}{\pi\sigma} \ln\!\left(\frac{|x|}{r_{0}}\right)\,,
\label{eq:variance}
\end{equation}
which has been verified in Monte Carlo simulations \cite{Muller:2004vv}.
Notice that the variance depends on a UV scale $r_0$.

Using \eqref{eq:DiracdeltaGaussian} in \eqref{eq:free energy in finite volume step 1} and using translation invariance along the wall, we obtain
\begin{equation}
    \Delta F\approx -A \lambda_0^2\int d^2rdwG\left(r,w+L_{z}\right)\dfrac{e^{-\dfrac{w^{2}}{2d^2(r)}}}{\sqrt{2\pi } d\left(r\right)}.\label{eq:to integrate numerically}
\end{equation}
Given that $mL_z\gg 1$,  we can expand the propagator
\begin{equation}
    G\left(r,w+L_{z}\right)=\dfrac{e^{-m|x|}}{4\pi|x|}\,
    ,\qquad |x| = \sqrt{r^2+(w+L_z)^2}= L_z + w+\frac{r^2+w^2}{2L_z}+\dots\,,
\end{equation}
leading to
\begin{align}
   \frac{\Delta F}{A}\approx -\lambda_{0}^{2}\dfrac{e^{-mL_{z}}}{4\pi L_{z}}\int dwe^{-mw}\int d^{2}re^{-\frac{mr^{2}}{2L_{z}}}\dfrac{e^{-\dfrac{w^{2}}{2d^{2}(r)}}}{\sqrt{2\pi}d\left(r\right)}.
\end{align}
Notice that the integral is dominated by $r^2\sim  L_z/m$, which leads to $w^2\sim d^2(r)\sim \frac{1}{\sigma}\ln L_z$. This is consistent with our approximations that used 
$|r|, w \ll L_z $ and $m w^2 \ll L_z$.
Integrating over $w$, yields 
\begin{align}
   \frac{\Delta F}{A}\approx - \lambda_0^2 \dfrac{e^{-mL_{z}}}{4\pi L_{z}}
   \int  d^2r \,e^{-\frac{mr^{2}}{2L_{z}}}
   e^{\frac{1}{2}m^2d^{2}(r)}\,.
\end{align}
Then, using the variance \eqref{eq:variance},
the integral over $r$ can be done exactly,
\begin{align}
    \int d^2r  
 e^{ \frac{1}{2}m^2d^2(r)-\frac{mr^2}{2L_z}}=
 2\pi \int_{r_0}^\infty dr r (r/r_0)^{2\chi}
 e^{-\frac{mr^2}{2L_z}}
\approx 
 2\pi (L_z/m)^{1+\chi} r_0^{-2\chi} 2^\chi \Gamma(1+\chi) \,,
\end{align}
where $\chi\equiv\frac{m^{2}}{4\pi\sigma}$ and we used $r_0$ as a UV cutoff for the integral over $r$. This is equivalent to integrating only over the region that has $d^2(r)>0$.
Finally, we can write
\begin{align}
\frac{\sigma_{\text{eff}}(L_z)-\sigma}{\sigma}
= \lim_{A\to \infty}\frac{\Delta F}{\sigma A}\approx
 - \lambda^2 \frac{2^\chi \Gamma(\chi)}{8\pi} (mL_z)^{\chi} e^{-mL_{z}}\,,\qquad mL_z\gg 1\,,
\label{eq:sigmaeffasymp}
\end{align}
where we introduced the renormalized dimensionless coupling
\begin{equation}
    \lambda^2\equiv \lambda_0^2\dfrac{m}{\sigma^2} \frac{1 }{(mr_0)^{2\chi}}
    \,.
    \label{renormalized lambda}
\end{equation}
As we shall see, this renormalized coupling will show up in other physical observables.

\begin{figure}
\centering\includegraphics[width=0.9\linewidth]{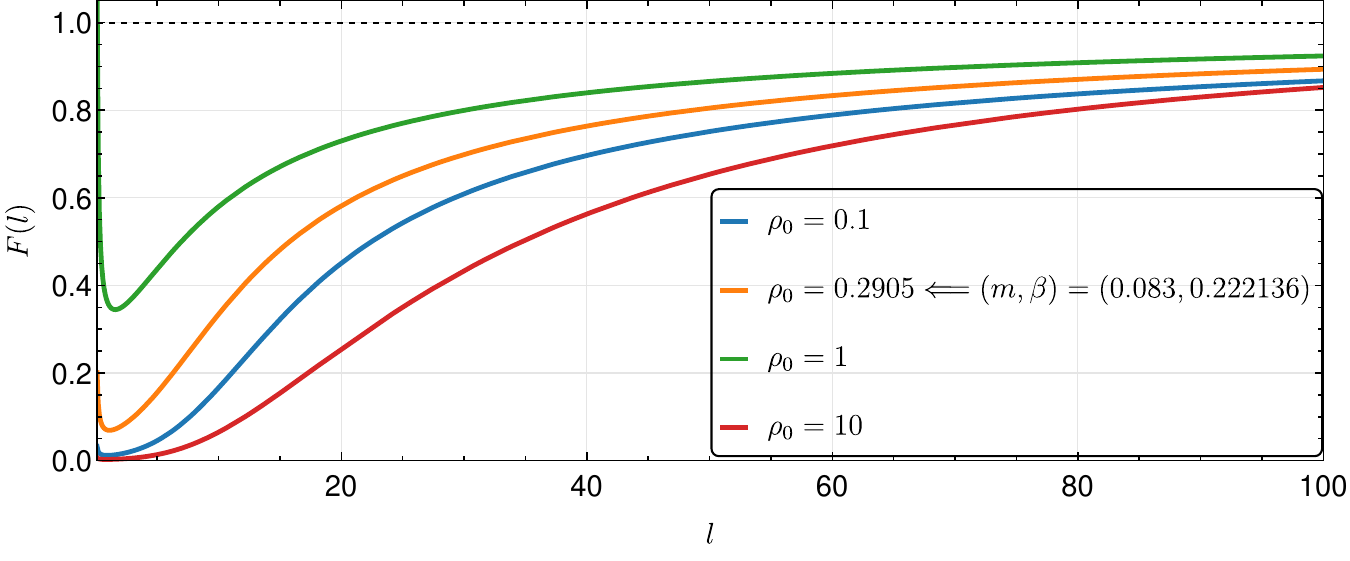}
    \caption{Numerical evaluation of $F(l)$ defined in eq.~\eqref{eq:definition F(l)} for $\rho_0=0.1,\,0.2905,\,1,\,10$. Both axes are truncated. The value $\rho_0=0.2905$ corresponds to the estimate of $r_0$ in eq.~\eqref{eq:r0} in sec.~\ref{sec:numeric}; at the corresponding temperature, $m a=0.083(4)$ with $a$ the lattice spacing.}
    \label{fig:f(l)}
\end{figure}

Equation \eqref{eq:sigmaeffasymp} gives the leading asymptotic behavior of the string tension when $mL_z \to \infty$. There are  corrections to this asymptotic result which we encode in a function $F$ defined by
\begin{align}
  \frac{\sigma_{\text{eff}}(L_z)-\sigma}{\sigma}
\approx
 - \lambda^2 \frac{2^\chi \Gamma(\chi)}{8\pi} (mL_z)^{\chi} e^{-mL_{z}}
 F(mL_z)
 \,.
 \label{eq:sigmaeffcorrected}
\end{align}
By construction $\lim_{l\to \infty}F(l)=1$ but the function $F$ approaches this asymptotic value with power-law corrections in inverse powers of $l$.
Some of these corrections are encoded in \eqref{eq:to integrate numerically} and give
\begin{align}
     F(l)= \frac{\rho_0^{2\chi}}{\Gamma(1+\chi) (2l)^\chi}
    \int_{-\infty}^\infty dw \frac{e^{-\frac{w^2}{2}}}{\sqrt{2\pi}} \int_{\rho_0}^\infty d\rho \frac{\rho}{s}e^{l-s}\,,\qquad
    s^2=  \rho^2+\left(l + w \sqrt{4\chi \ln\frac{\rho}{\rho_0}}\right)^2\,,\label{eq:definition F(l)}
\end{align}
with $\rho_0 = mr_0$. 
The plot of $F(l)$ in fig.~\ref{fig:f(l)} shows that the asymptotic behavior $F(l)\approx 1$ only sets in for rather large  $l=mL_z$.

There are other terms in the effective action, like for example $\lambda_k\int d^2x \sqrt{h} \,\nabla_\perp^k\phi(x,\pi(x))$, where $\nabla_\perp=n^\mu
\partial_\mu$ and $n^\mu$ a unit vector normal to the domain wall.
Notice that $\nabla^2=\nabla_\perp^2+\nabla_\parallel^2=m^2$ on-shell. This means that we only need to consider $k=0$ and $k=1$.
As explained in \eqref{eq:glueballmomentum}, we have $\nabla_\perp\approx \pm i m$,
therefore, both choices lead to the same leading behavior, but  differ at subleading orders in the expansion at large $L_z$. 
Therefore, the subleading terms depend on another Wilson coefficient and are not completely determined by the coupling $\lambda$.\footnote{
Strictly speaking,  $\lambda^2$ in \eqref{eq:sigmaeffasymp} is a renormalized version of the combination $\lambda_0^2+m^2\lambda_1^2$. }

The effective field theory may also contain an interaction of the form 
\begin{equation}\label{eq:quadratic interaction}
    \lambda_2\int d^2x \sqrt{h}  \phi^2(x,\pi(x))\,.
\end{equation}
This gives the following contribution to the free energy of the wall (see fig. \ref{fig:cubic diagrams})
\begin{align}
\Delta F &\approx -
 \lambda_2 \int d^2x   \,
\Big[
\big\langle \phi(x ,\pi(x ))\,\phi(x ,\pi(x ))\big\rangle_{L_z}
-
\big\langle \phi(x ,\pi(x ))\,\phi(x ,\pi(x ))\big\rangle_{\infty}
\Big]
\\
&\approx  -2 \lambda_2 A G(0,0,L_z) = -A \frac{\lambda_2}{2\pi L_z} e^{-mL_z} \,.
\end{align}
We see that this is subleading with respect to \eqref{eq:sigmaeffasymp} at large $L
_z$ (although it has the same exponential dependence).

\subsection{Particle scattering across the string }\label{sec:s-matrix}

Consider a static infinite straight string in 3D Minkowski spacetime. The worldsheet of the string is a $\mathbb{M}^2$ surface embedded in $\mathbb{M}^3$.
Consider now a particle of mass $m$ in an incoming plane wave state of momentum $p=(p_\parallel,p_\perp)$ with $p_\parallel \in \mathbb{M}^2$ the momentum along the string worldsheet and $p_\perp \in \mathbb{R}$ the momentum transverse to it.
For an on-shell particle, we have $m^2=-p_\perp^2 -p_\parallel^2$.
Therefore, $s\equiv -p_\parallel^2$  determines the kinematics of the initial state.
We will be interested in the transition amplitude to a final state with a string without internal excitations and a single bulk particle moving away from it.
Given that the parallel momentum $p_\parallel$ is conserved and that the final particle is on-shell, we conclude that the final $p_\perp$ is either the same or  minus the initial $p_\perp$.
In other words, the particle can scatter across the string and keep its momentum or reflect on the string and then $p_\perp\to -p_\perp$. 

Physical scattering requires $s=m^2+p_\perp^2 > m^2$, which is beyond the range of validity of EFT.
However, the scattering amplitude $\mathcal{M}(s)$ is an analytic function of $s$ and the regime $s\ll m^2$ is controlled by our EFT. 

To leading order in the interaction \eqref{eq:interaction}, the scattering amplitude is given by 
\begin{align}
\lambda_0^{2}\int d^{2}x_1 \,
d^{2}x_2 \,
\Big\langle 
e^{i p_{\parallel}^\text{in}\cdot x_1+ ip_{\perp}^\text{in}\,\pi(x_1)}\,
e^{-i p_{\parallel}^\text{out}\cdot x_2 -ip_{\perp}^\text{out}\,\pi(x_2)}
\Big\rangle\,.
\end{align}
Writing $x_2=x_1+x$ and $\pi(x_2)=\pi(x_1+x)-\pi(x_1)+\pi_0$, 
where we identified $\pi(x_1)$ with the zero-mode $\pi_0$, we can integrate over $x_1$ to obtain 
\begin{align}
(2\pi)^2\delta^2(p_{\parallel}^\text{in}-p_{\parallel}^\text{out})\lambda_0^{2}\int 
d^{2}x \,
\Big\langle 
e^{ ip_{\perp}^\text{in}\,\pi_0}\,
e^{-i p_{\parallel}^\text{out}\cdot x -ip_{\perp}^\text{out}(\pi(x_1+x)-\pi(x_1)+\pi_0)}
\Big\rangle\,.
\end{align}
Notice that $\pi(x_1+x)-\pi(x_1)$ is a random variable that depends only on $x$, so the integrand is independent of $x_1$ as we assumed.
From the conservation of $p_\parallel$ and the on-shellness condition, we conclude that 
$p_{\perp}^\text{out}=\pm p_{\perp}^\text{in}$ as expected. 

Let us first consider the transmission amplitude of a particle across a string  ($p_{\perp}^\text{out}=p_{\perp}^\text{in}$). In this case, we can write the reduced amplitude as
\begin{align}
\mathcal{T}(s)
&\approx\lambda_0^{2}\int d^{2}x \,
e^{-i p_{\parallel}\cdot x}\,
\Big\langle e^{-i p_{\perp}(\pi(x)-\pi(0))}\Big\rangle
\approx\lambda_0^{2}\int d^{2}x \,
e^{-i p_{\parallel}\cdot x}\,e^{-\frac{1}{2}p_\perp^2 d^2(x)}
\\
&\approx\lambda_0^{2}\int d^{2}x \,
e^{-i p_{\parallel}\cdot x}
\left(\frac{|x|}{r_{0}}\right)^{2\chi}
\approx\lambda_0^{2} r_0^{-2\chi}
\frac{\pi  
    \Gamma
   (\chi +1)}{\Gamma (-\chi )}\left(\frac{4}{p_\parallel^2}\right)^{\chi +1}\\
   &\approx 
 -\lambda^{2} 
 \frac{m}{16\pi}
\frac{\Gamma(\chi)}{\Gamma(1-\chi)}
\left(\frac{-s}{4m^2}  \right)^{-\chi-1}\,,\qquad |s|\ll m^2\,.
\label{eq:MrE_def}
\end{align}
Again, the final result can be expressed in terms of the renormalized coupling  \eqref{renormalized lambda}.
Our effective coupling makes a prediction for the behavior of the analytically continued amplitude at the beginning of the cut for $s>0$.
The discontinuity of this cut is related to decays of the bulk particle into massless branon modes on the string.

Let us now comment on the reflection amplitude. The reduced amplitude reads
\begin{align}
\mathcal{R}(s)
&\approx\lambda_0^{2}\int d^{2}x \,
e^{-i p_{\parallel}\cdot x}\,
\Big\langle e^{2ip_\perp \pi_0}e^{i p_{\perp}(\pi(x)-\pi(0))}\Big\rangle
\approx 
\Big\langle e^{2ip_\perp \pi_0}
\Big\rangle
\mathcal{T}(s)\,.
\end{align}
This means that the reflected amplitude depends on the state of the transverse zero mode of the string. For example, if $\pi_0$ is localized then $|\mathcal{R}|=|\mathcal{T}|$, while if $\pi_0$ is completely delocalized, then $\mathcal{R}=0$.

\subsection{Correlation functions in the presence of a wall}\label{sec:correlation functions}
Let us first consider the {\bf one-point} function in the presence of the wall.
At leading order in the coupling $\lambda_0$, the difference between anti-periodic and periodic boundary conditions is
\begin{align}
\delta\langle \phi \rangle \equiv \langle \phi\rangle_{\text{AP}}-\langle \phi\rangle_{\text{P}}
&=
\lambda_0 
\int d^{2}x 
\big\langle \phi(0)\,\phi(x,\pi(x))\big\rangle
\nonumber\\
&=
\lambda_0
\int d^{2}x \int ds \rho(s)
\int\frac{d^{3}k}{(2\pi)^{3}}\;
\frac{e^{ik_{\parallel}\cdot x}}{k^{2}+s}\,
\Big\langle e^{ik_{\perp}\pi(x)}\Big\rangle.
\label{eq:1pt_avg_start}
\end{align}
The average over the zero mode $\pi_0$ of $\pi(x)$ leads to 
\begin{equation}
\Big\langle e^{ik_{\perp}\pi(x)}\Big\rangle=\frac{1}{L_z}\int_0^{L_z} d\pi_0 \,e^{-ik_\perp \pi_0 } \Big\langle e^{ik_{\perp}\tilde{\pi}(x)}\Big\rangle
=\frac{2\pi}{L_z}\,\delta(k_\perp)\Big\langle e^{ik_{\perp}\tilde{\pi}(x)}\Big\rangle
=\frac{2\pi}{L_z}\,\delta(k_\perp)
\,,
\end{equation}
where we used $\pi(x)=\pi_0+\tilde{\pi}(x)$.
 Performing the remaining $d^2x$ integral then sets $k_\parallel=0$, yielding
\begin{align}\label{eq:delta epsilon EFT}
\delta\langle \phi\rangle
&=
\frac{\lambda_0 }{L_{z} }\int\frac{ds}{s}\rho(s)\,.
\end{align}
Thus, we conclude that the one-point function depends on the full spectral density of the operator, unlike the observables considered above that were dominated by the delta-function $\delta(s-m^2)$ corresponding to an on-shell single-particle.

\begin{figure}[t]
    \centering
    \includegraphics[width=0.49\linewidth]{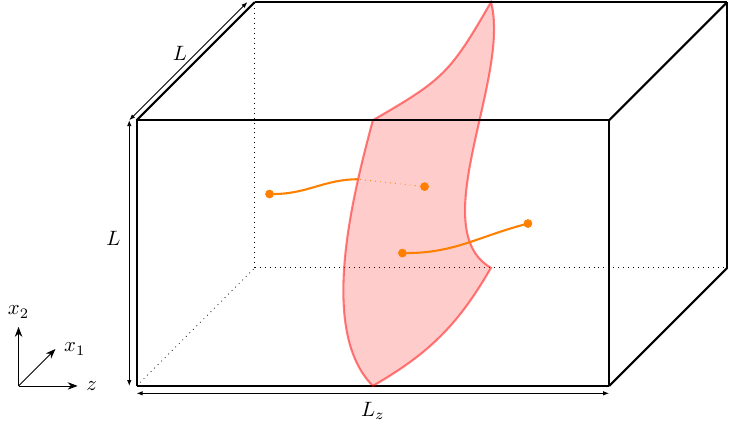}    \includegraphics[width=0.49\linewidth]{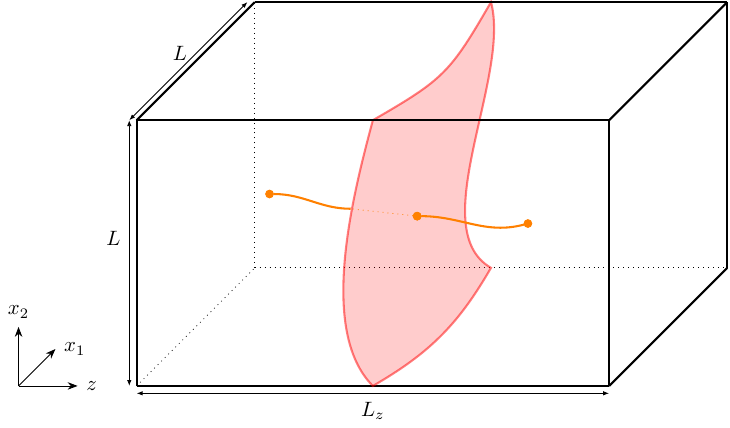}
    \caption{Left) diagrams contributing at $\mathcal{O}(\lambda^2)$ to the correction to the 2pt function for the linear coupling in eq.~\eqref{eq:interaction}; right) diagrams contributing at $\mathcal{O}(\lambda_2)$  to the correction to the 2pt function for the quadratic coupling in eq.~\eqref{eq:quadratic interaction}.}
    \label{fig:2pt connected}
\end{figure}

We next consider the connected \textbf{two-point} function in the wall background. The leading correction
involves two insertions of the interaction vertex (see Fig.~\ref{fig:2pt connected}) and takes the form
\begin{equation}
\delta\langle\phi(0)\phi(x)\rangle\approx\frac{\lambda_0^2 }{2}\int d^2{x }_1\,d^2{x }_2\;
\big\langle
\phi(0)\,\phi({x }_1,\pi({x }_1))\,
\phi({x }_2,\pi({x }_2))\,\phi(x_\parallel,x_\perp)
\big\rangle\,,
\end{equation}
where $\delta \langle\dots\rangle = \langle\dots\rangle_{\rm AP}^\text{conn} - \langle\dots\rangle_{\rm P}^\text{conn} $.
Using Wick's theorem for the bulk field and keeping only the cross-contractions (linking $0$ and $x$ to the worldsheet insertions), the two such contractions cancel the explicit factor of $1/2$. Writing the $\langle \phi \phi\rangle$ propagator in Fourier space and using $k=(k_{\parallel},k_{\perp})$, we obtain
\begin{align}
\delta\langle\phi(0)\phi(x)\rangle
&\approx
 \lambda_0^{2} \int d^{2}x_{1}\,d^{2}x_{2} \int ds_1\rho(s_1) \int ds_2\rho(s_2)
\\
&\times \int\frac{d^{3}k}{(2\pi)^{3}}\frac{e^{ik_{\parallel}\cdot x_{1}}}{k^{2}+s_1}
\int\frac{d^{3}k'}{(2\pi)^{3}}\frac{e^{ik'_{\parallel}\cdot(x_{2}-x_{\parallel})-ik'_{\perp}x_{\perp}}}{(k')^{2}+s_2}
\nonumber
\Big\langle e^{ik_{\perp}\pi(x_{1})}\,e^{ik'_{\perp}\pi(x_{2})}\Big\rangle .
\label{eq:deriv_line1}
\end{align}
The branon correlator gives
\begin{align}
    \Big\langle e^{ik_{\perp}\pi(x_{1})}\,e^{ik'_{\perp}\pi(x_{2})}\Big\rangle =
    \frac{2\pi}{L_{z}}\delta(k_{\perp}+k'_{\perp})\Big\langle e^{ik_{\perp}\left[\pi(x_{1})-\pi(x_{2})\right]}\Big\rangle=
    \frac{2\pi}{L_{z}}\delta(k_{\perp}+k'_{\perp})e^{-\frac{1}{2}k_\perp^2d^2(x_1-x_2)}\,,
\end{align}
where the delta-function follows from averaging over the zero mode of $\pi$.
It is now convenient to write $x_2=x_1 +y$ and change to the integration variables $y$ and $x_1$. Integrating over $x_1$ produces a delta function enforcing the conservation of parallel momentum,
\begin{equation}
\int d^{2}x_1\,e^{i(k_{\parallel}+k'_{\parallel})\cdot x_1}=(2\pi)^{2}\delta^{2}(k_{\parallel}+k'_{\parallel})\,.
\end{equation}
Thus, we find
\begin{align}
\delta\langle\phi(0)\phi(x)\rangle
&\approx
 \frac{\lambda_0^{2}}{L_z} \int d^{2}y \int ds_1\rho(s_1) \int ds_2\rho(s_2)
\\
&\times \int\frac{d^{3}k}{(2\pi)^{3}}\frac{1}{k^{2}+s_1}
\frac{e^{-ik_{\parallel}\cdot(y-x_{\parallel})+ik_{\perp}x_{\perp}}}{k^{2}+s_2}
e^{-\frac{1}{2}k_\perp^2d^2(y)}\,.
\nonumber
\end{align}

Let us now consider the integral over $d^2k_\parallel$. Recall that the denominator $k^2+s_1 = k^2_\parallel + k^2_\perp+s_1$ and $k^2_\perp+s_1 >m^2$. Therefore, integrating over $k_\parallel$ produces a function of $(y-x_\parallel)^2$ that localizes 
$(y-x_\parallel)^2 \lesssim 1/m^2$. In practice, we can obtain the leading behavior of this approximation simply by neglecting $k_\parallel$ in the denominators and using
\begin{align}
    \int \frac{d^{2}k_\parallel}{(2\pi)^2}
    e^{-ik_{\parallel}\cdot(y-x_{\parallel})}=\delta^2(y-x_{\parallel})\,.
\end{align}
This leads to
\begin{align}
\delta\langle\phi(0)\phi(x)\rangle
&\approx
 \frac{\lambda_0^{2}}{L_z}  \int ds_1\rho(s_1) \int ds_2\rho(s_2)
 \int\frac{dk_\perp}{2\pi}
\frac{e^{ik_{\perp}x_{\perp}-\frac{1}{2}k_\perp^2d^2(x_\parallel)}}{(k_\perp^{2}+s_1)(k_\perp^{2}+s_2)}\,.
\end{align}

This formula has two interesting regimes. Firstly, if $m d(x_\parallel)\gg 1$ and $x_\perp$ is not too large, then we can neglect the $k_\perp$ dependence in the denominator and obtain
\begin{align}
\delta\langle\phi(0)\phi(x)\rangle
&\approx
 \frac{\lambda_0^{2}}{L_z}  \left[\int \frac{ds}{s}\rho(s)\right]^2 
 \frac{1}{\sqrt{2\pi d^2(x_{\parallel})}}\,
\exp\!\left[-\frac{x_{\perp}^{2}}{2 d^2(x_{\parallel})}\right]
\,.
\label{eq:2ptfunctionGaussianprediction}
\end{align}
This regime is not universal because it depends on the full spectral density of the correlator. However, for a given operator,  the ratio
\begin{align}
\frac{
\delta\langle\phi(0)\phi(x)\rangle}{\left[ \delta\langle\phi\rangle\right]^2} 
&\approx
 \frac{L_z}{\sqrt{2\pi d^2(x_{\parallel})}}\,
\exp\!\left[-\frac{x_{\perp}^{2}}{2 d^2(x_{\parallel})}\right]
\,,
\label{eq:2ptover1ptsquared}
\end{align}
is given by a universal Gaussian in $x_\perp$ with variance given by $d^2(x_\parallel)$.

Secondly, in the regime $m d(x_\parallel)$ is not too large and $1/m\ll |x_\perp|<  L_z/2$, the integral over $k_\perp$ is dominated by the pole at $k_\perp^2=-m^2$ and leads to
\begin{align}
\delta\langle\phi(0)\phi(x)\rangle
&\approx\frac{\lambda_0^{2} }{L_{z}} \int\frac{dk_{\perp}}{2\pi}\frac{e^{ ik_{\perp}x_{\perp}}}{\left( k_{\perp}^{2}+m^{2}\right)^{2}}
e^{-\frac{1}{2}k_\perp^2d^2(x_\parallel)}
\\
&\approx\frac{\lambda_0^{2} }{4m^2L_{z}}  e^{ -m|x_{\perp}|} 
e^{\frac{1}{2}m^2d^2(x_\parallel)}
|x_\perp| \,.
\label{eq:2ptlambda0exponential}
\end{align}
The validity of this approximation requires that $m|x_\perp|\gg m^2 d^2(x_\parallel)$ so that the second exponential is a small correction to the first one. Notice that this is equivalent to $m|x_\perp|\ll \frac{x_\perp^2}{ 2d^2(x_\parallel)}$ which means that this approximation is valid when the prediction from \eqref{eq:2ptfunctionGaussianprediction} is (exponentially) smaller.
Finally, we can write
\begin{align}
\delta\langle\phi(0)\phi(x)\rangle
&\approx \frac{\lambda^{2} m }{64\pi^2 \chi^2}  
\left( m|x_\parallel|\right)^{2\chi}
\frac{|x_\perp|}{L_z} e^{ -m|x_{\perp}|} \,, \qquad |x_\perp|\gg m d^2(x_\parallel)\gg \frac{1}{m}\,,
\end{align}
where we used the effective coupling
\eqref{renormalized lambda}.
This is the third observable that is controlled by the coupling $\lambda$. 

It is instructive to rederive the last formula without using Fourier space. We start from
\begin{equation}
\delta\langle\phi(0)\phi(x)\rangle\approx \lambda_0^2  \int d^2{x }_1\,d^2{x }_2\;
\left\langle
\frac{e^{-m\sqrt{x_1^2+\pi(x_1)^2}}}{4\pi\sqrt{x_1^2+\pi(x_1)^2}}
\frac{e^{-m\sqrt{(x_2-x_\parallel)^2+(x_\perp-\pi(x_2))^2}}}{4\pi\sqrt{(x_2-x_\parallel)^2+(x_\perp-\pi(x_2))^2}
}\right\rangle\,,
\end{equation}
where we assumed that $0<\pi(x_1),\pi(x_2)<x_\perp$
and that the $\phi$ propagator is dominated by its large distance behavior. This is true for $m\pi(x_1)\gg 1 $ and $m(x_\perp-\pi(x_1))\gg 1 $.
In this case, the integral over $x_1$ is dominated by small values of $x_1$ and the integral over $x_2$ is dominated by $x_2$ close to $x_\parallel$.
Therefore, we expand the square roots to the leading order in the denominator and keep a subleading term in the exponent,
\begin{equation}
\delta\langle\phi(0)\phi(x)\rangle\approx \lambda_0^2  \int d^2{x }_1\,d^2{x }_2\;
\Big\langle
\frac{e^{-m \pi(x_1) -m\frac{x_1^2}{2\pi(x_1)}}}{4\pi\, \pi(x_1)}
\frac{e^{-m(x_\perp-\pi(x_2))  -m\frac{(x_2-x_\parallel)^2}{2(x_\perp-\pi(x_2))}}}{4\pi (x_\perp-\pi(x_2))}
\Big\rangle\,.
\end{equation}
Now we perform the gaussian integrals over $x_1$ and $x_2$ assuming that $\pi(x_1)\approx \pi(0)$ and $\pi(x_2)\approx \pi(x_\parallel)$. This leads to 
\begin{align}
\delta\langle\phi(0)\phi(x)\rangle \approx \frac{\lambda_0^2 }{4m^2} e^{-m x_\perp}
 \langle
e^{-m (\pi(0) -\pi(x_\parallel))}
 \rangle 
 \approx \frac{\lambda_0^2 }{4m^2} e^{-m x_\perp}
e^{\frac{1}{2}m^2 d^2 (x_\parallel)} \frac{x_\perp}{L_z}\,,
\end{align}
where we included the factor $x_\perp/L_z$ that represents the probability of the zero mode being between 0 and $x_\perp$ as we assumed. This matches precisely \eqref{eq:2ptlambda0exponential} for $x_\perp>0$.

\section{Geometric approximation for correlation functions}

\label{sec:theory-geom}

Sec.~\ref{sec:theory} formulated the problem in terms of an effective coupling between the fluctuating wall and a bulk field creating the lightest massive particle. In the present section we discuss a complementary thin-wall description in which the wall is treated directly as a fluctuating interface separating the two vacua. The purpose of this section is not to derive the EFT couplings microscopically, but to isolate those features of the correlators that follow already from rough-wall kinematics together with a short-distance wall-core profile. This provides an independent interpretation of the nearby-regime results and clarifies which statements do not rely on a quantitative extraction of $\lambda$.

\subsection{\texorpdfstring{$\mathbb{Z}_2$}{Z2}-odd operators}
\label{sec:theory-z2-odd-operators}

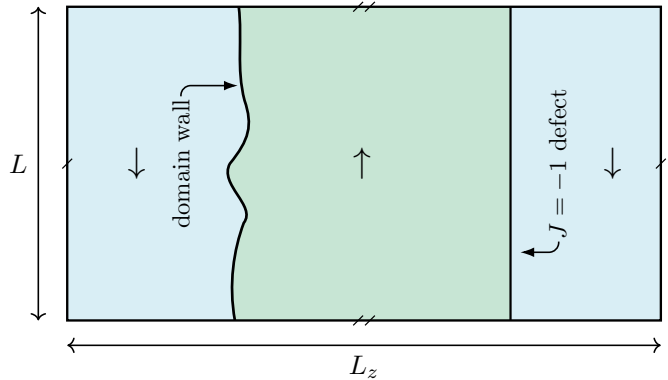
\begin{figure}[t]
  \centering
  \begin{tikzpicture}[x=0.55cm,y=0.55cm,line cap=round,line join=round]
	\begin{pgfonlayer}{nodelayer}
		\node [style=none] (A) at (0,0) {};
		\node [style=none] (B) at (0,8.2) {};
		\node [style=none] (C) at (15.4,8.2) {};
		\node [style=none] (D) at (15.4,0) {};

		\node [style=none] (E) at (11.5,0) {};
		\node [style=none] (F) at (11.5,8.2) {};

		\node [style=none] (W0) at (4.35,0.00) {};
		\node [style=none] (W1) at (4.58,2.55) {};
		\node [style=none] (W2) at (4.22,4.10) {};
		\node [style=none] (W3) at (4.60,5.75) {};
		\node [style=none] (W4) at (4.45,8.20) {};

		\node [style=label text, rotate=90] (DWlab) at (2.95,4.10) {domain wall};
		\node [style=label text, rotate=90] (Jlab) at (12.75,4.10) {$J=-1$ defect};

		\node [style=spin text] (spinL) at (1.80,4.1) {$\downarrow$};
		\node [style=spin text] (spinM) at (7.65,4.1) {$\uparrow$};
		\node [style=spin text] (spinR) at (14.15,4.1) {$\downarrow$};

		\node [style=none] (LzL) at (0,-0.65) {};
		\node [style=none] (LzR) at (15.4,-0.65) {};
		\node [style=none, label={below:$L_z$}] (Lzlab) at (7.7,-0.65) {};
		\node [style=none] (LLB) at (-0.75,0) {};
		\node [style=none] (LLT) at (-0.75,8.2) {};
		\node [style=none, label={left:$L$}] (Llab) at (-0.75,4.1) {};
	\end{pgfonlayer}

	\begin{pgfonlayer}{edgelayer}
		\fill[white] (-1.25,-1.15) rectangle (16.15,8.9);

		\fill[cyanregion] (A.center) rectangle (C.center);

		\fill[greenregion]
			(W0.center)
			.. controls (4.18,0.95) and (4.32,1.85) .. (W1.center)
			.. controls (4.92,3.00) and (3.92,3.55) .. (W2.center)
			.. controls (4.72,4.72) and (4.82,5.10) .. (W3.center)
			.. controls (4.38,6.60) and (4.55,7.35) .. (W4.center)
			-- (F.center) -- (E.center) -- cycle;

		\draw[main edge] (A.center) -- (B.center) -- (C.center) -- (D.center) -- cycle;
		\draw[separator edge] (E.center) -- (F.center);
		\draw[domain wall edge]
			(W0.center)
			.. controls (4.18,0.95) and (4.32,1.85) .. (W1.center)
			.. controls (4.92,3.00) and (3.92,3.55) .. (W2.center)
			.. controls (4.72,4.72) and (4.82,5.10) .. (W3.center)
			.. controls (4.38,6.60) and (4.55,7.35) .. (W4.center);
        \draw[ glued one] (A) -- (B);    %
        \draw[ glued one] (D) -- (C);  %
        
        \draw[ glued two] (B) -- (C);   %
        \draw[ glued two] (A) -- (D);   %
		\draw[annotation edge, rounded corners=4pt]
    (DWlab.east) -- ++(0,0.28) -- ++(1.48,0.0);

        \draw[annotation edge, rounded corners=4pt]
    (Jlab.west) -- ++(0,-0.28) -- ++(-1,0.0);

		\draw[dimension edge] (LzL.center) -- (LzR.center);
		\draw[dimension edge] (LLB.center) -- (LLT.center);
	\end{pgfonlayer}
\end{tikzpicture}
  \caption{Schematic representation of a cut at fixed $x$ (equivalently, at fixed $y$). Due to the periodic boundary conditions, the spins change sign at two locations: at the $J=-1$ defect and at the domain wall. In the text, as is customary, we will often refer to the former simply as the anti-periodic boundary condition, although here it is shown explicitly. The figure depicts only one possible simplified spin configuration, since all spins could be reversed, and clusters, which are typically present near the critical point, have been omitted for clarity. Only the domain wall carries a local excess of energy.}
  \label{fig:wall+defect}
\end{figure}
We begin with a $\mathbb{Z}_2$-odd bulk operator, for concreteness the spin field $s$.
Far away from the wall, $s$ approaches $\pm \langle s\rangle$, with $\langle s\rangle$ the spontaneous magnetization.
However, we have $\langle s\rangle_{\text{AP}} =0$ because it is equally likely that a fixed spin will be on either side of the fluctuating wall.

There are two surfaces at which $s$ changes sign: the (topological) defect and the domain wall -- see figure \ref{fig:wall+defect}. To isolate the domain wall effects, we use a compensating $\mathbb{Z}_2$ line between the spins, 
\begin{equation}
G_{s}^{\rm def}(x)
\;\equiv\;
\Big\langle s(0)\,\eta(x,0)\,s(x)\Big\rangle_{\rm AP},
\qquad
\eta(x,0)=\pm1\,,
\label{eq:s-eta-s}
\end{equation}
with $\eta(x,y)=-1$ iff the shortest path from $x$ to $y$ crosses the defect plane. 
At long distances, this  correlator reduces to $\langle s\rangle^{2}$ times a $\mathbb{Z}_2$ sign that records whether the two insertions lie in the same magnetized domain for a given wall configuration.  It is convenient to perform the  average over $\pi(0)$ first, with fixed relative wall displacement $w\equiv \pi(x_\parallel)-\pi(0)$.
This average over the uniformly distributed variable $\pi(0)$ yields the expectation value of the sign,
\begin{equation}
\big\langle \mathrm{sgn}\!\big(x_\perp-\pi(x_\parallel)\big)\,
\mathrm{sgn}\!\big(-\pi(0)\big)\big\rangle_{w}
\;=\;\frac{L_z-2|w-x_\perp|}{L_z}\,,
\qquad (|w-x_\perp|\le L_z/2),
\end{equation}
where  the restriction implements the shortest-separation convention.
As discussed above, $w$ is Gaussian with variance $d^{2}(x_\parallel)$, therefore
\begin{align}
 G_{s}^{\rm def}(x)
&=
\langle s\rangle^2
\int dw\;
\frac{e^{-\frac{w^2}{2d^2(x_\parallel)}}}{\sqrt{2\pi d^2(x_\parallel)}}\,
\frac{L_z-2|w-x_\perp|}{L_z}\,.
\label{eq:2pt-odd-general}
\end{align}

 Evaluating this in the limit $L_z\gg d(L/2)$, so that the integration domain can be extended to $\mathbb{R}$, yields the universal form
\begin{equation}
\frac{ G_{s}^{\rm def}(x)}{\langle s\rangle^2}
\approx
1-\frac{2d(x_\parallel)}{L_z}\,f\!\left(\frac{|x_\perp|}{d(x_\parallel)}\right),
\label{eq:2pt-odd-univ}
\end{equation}
with
\begin{equation}
f(y)=
y\,\mathrm{erf}\!\left(\frac{y}{\sqrt{2}}\right)
+\sqrt{\frac{2}{\pi}}\,e^{-y^2/2}\,.
\end{equation}

This holds in the regime $L_z\gg |x_\perp|$ and $d(x_\parallel)\gg 1/m=\xi $, which   is the scale of the wall thickness. %

This computation can be interpreted as a prediction of the EFT of the previous section. We treated the domain wall as an infinitely thin fluctuating surface with action \eqref{eq:EST-free} and considered its effect on a $\mathbb{Z}_2$-odd field whose one-point function has opposite sign 
on the two sides of the wall.

\subsection{\texorpdfstring{$\mathbb{Z}_2$}{Z2}-even operators}

We now turn to $\mathbb{Z}_2$-even operators, for which the defect does not induce any sign change. As a representative even operator we take the lattice energy density $\epsilon \sim s_i s_j J_{ij}$.

In the thin-wall picture, the dominant effect on even operators is localized in the wall core: after subtracting the periodic bulk value, $\epsilon-\langle\epsilon\rangle_{\rm P}$ has support only near the wall core. We parametrize this by a transverse profile $p_\epsilon$. For a fixed wall configuration, we define
\begin{equation}
p_\epsilon\!\left(x_\perp - \pi(x_\parallel) \right)
\equiv
\langle \epsilon(x)\rangle_{\pi} - \langle \epsilon\rangle_{\rm P}\,,
\label{eq:peps_def}
\end{equation}
where we neglected the curvature of the wall.
A convenient integrated measure of the wall-core strength is
\begin{equation}
\mathcal{T}_\epsilon \equiv \int dx_\perp\, p_\epsilon(x_\perp)\,.
\label{eq:T_def}
\end{equation}

By translational invariance, the difference of expectation values, $\delta\langle \epsilon\rangle= \langle \epsilon(x)\rangle_{\rm AP}-\langle \epsilon\rangle_{\rm P}
$, is equal to its volume average.
Therefore,
\begin{equation}
\delta\langle \epsilon\rangle
\;\equiv\;
\int \frac{dx_\perp}{L_z} \left[ \langle \epsilon(x)\rangle_{\rm AP}-\langle \epsilon\rangle_{\rm P}
\right]
=
\frac{1}{L_z}\int dx_\perp\,p_\epsilon(x_\perp-\pi(x_\parallel))
=
\frac{\mathcal{T}_\epsilon}{L_z}\,.
\label{eq:eps_1pt_ToverLz}
\end{equation}
Thus the $1/L_z$ scaling is purely kinematic: it reflects the zero-mode delocalization, while all microscopic information is contained in $\mathcal{T}_\epsilon$.  
This matches the prediction \eqref{eq:delta epsilon EFT}.

For two-point functions, we consider the domain wall-induced difference
\begin{equation}
\delta G_\epsilon(x)\equiv
\langle \epsilon(0)\epsilon(x)\rangle^{\rm conn}_{\rm AP}
-
\langle \epsilon(0)\epsilon(x)\rangle^{\rm conn}_{\rm P}\,.
\label{eq:DeltaGeps_def}
\end{equation}
In the thin-wall picture the leading connected contribution comes from configurations in which both operator insertions overlap with the wall core. Writing the relative wall displacement $w\equiv \pi(x_\parallel)-\pi(0)$, with variance $d^2(x_\parallel)=\langle w^2\rangle$, and using the Gaussian approximation for $w$ in the infrared, the averages over the field $\pi$ give
\begin{equation}
\delta G_\epsilon(x)=
\int \frac{dw}{\sqrt{2\pi d^2(x_\parallel)}}\,
e^{-w^2/(2d^2(x_\parallel))}
\int_0^{L_z}\frac{dz}{L_z}\,
p_\epsilon(-z)\,p_\epsilon  (x_\perp-z-w),
\label{eq:DeltaGeps_overlap_exact}
\end{equation}
where the integration variable $z$ denotes $\pi(0)$. 

Assuming wall wandering is large compared to its thickness, $d(x_\parallel)\gg \xi = m^{-1}$ and that the separation between operator insertions  is also larger than $\xi$, then the integral over $z$ is peaked around 0, while the integral over $w$ is peaked around $x_\perp$. More precisely,   
\begin{equation}
\delta G_\epsilon(x)
\simeq
\frac{1}{L_z}\,
\frac{e^{-x_\perp^2/(2d^2(x_\parallel))}}{\sqrt{2\pi d^2(x_\parallel)}}\,
\left(\int dz p_\epsilon(z)\right)^2
=
\frac{\mathcal{T}_\epsilon^2}{L_z}\,
\frac{e^{-x_\perp^2/(2d^2(x_\parallel))}}{\sqrt{2\pi d^2(x_\parallel)}}
\qquad
\big(L_z\gg d(x_\parallel)\gg \xi \big)\,.
\label{eq:DeltaGeps_gaussian}
\end{equation}
 Thus, the transverse dependence is approximately Gaussian, with width set by $d^2(x_\parallel)$ and overall normalization fixed by $\mathcal{T}_\epsilon$. This matches precisely the prediction \eqref{eq:2ptover1ptsquared}.

\section{Monte Carlo simulations of domain walls in  3D Ising}\label{sec:numeric}

This section presents the lattice setup and analysis used to test the predictions of sections  \ref{sec:theory} and \ref{sec:theory-geom}. We test the nearby-regime kinematic relation in eq.~\eqref{eq:2ptover1ptsquared}, extract the wall-variance profile from odd-sector correlators, and perform cross-sector checks. We also present tentative evidence for the long-distance behavior in eq.~\eqref{eq:2ptlambda0exponential}. Our data does not reach the asymptotic distances required for a direct determination of $\lambda$, and a quantitative test of the free-energy prediction is beyond current precision.

We consider the standard nearest-neighbor 3D Ising model with Hamiltonian
\begin{equation}
    H = - \sum_{\langle i,j \rangle} J_{ij} s_i s_j,
\end{equation}
where $J_{ij}=1$ on all bonds except those crossing the plane $z=0$. On that plane we take $J_{ij}=+1$ in the periodic sector and $J_{ij}=-1$ in the anti-periodic sector. We simulate lattices of volume $V=L^2\times L_z$ in the symmetry-broken phase ($T<T_c$) which has two magnetized vacua separated by domain walls. In the anti-periodic sector, we choose $L$ large enough that configurations are dominated by a single domain wall. Bulk masses are extracted from connected two-point functions measured in the periodic sector.

\begin{figure}[t]
    \centering
    \includegraphics[width=\linewidth]{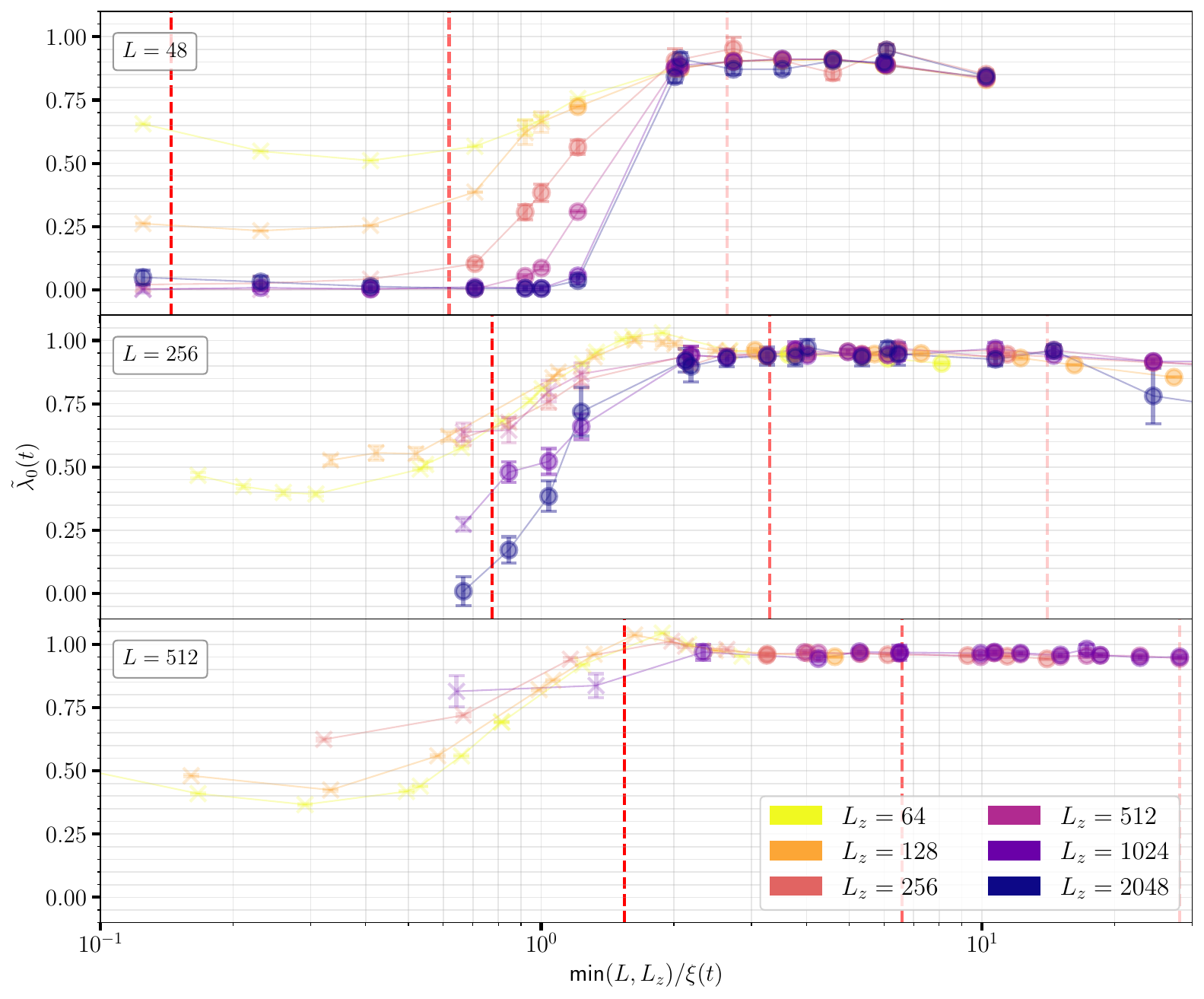}
    \caption{$\tilde\lambda_0$ extracted from the wall-induced one-point function, plotted against $\min(L,L_z)/\xi(t)$ for several domain-wall areas $A=L^2$ (panels) and transverse sizes $L_z$ (colors). Crosses mark data with $L_z<3\xi$, where transverse finite-size effects are expected. Vertical dashed lines indicate the reference locations corresponding to $t=10^{-4},10^{-3},10^{-2}$ under the identification $\min(L,L_z)=L$, and are included only to provide a rough sense of the distance from criticality.}
    \label{fig:1ptfunction z2 even}
     \centering
    \includegraphics[width=\linewidth]{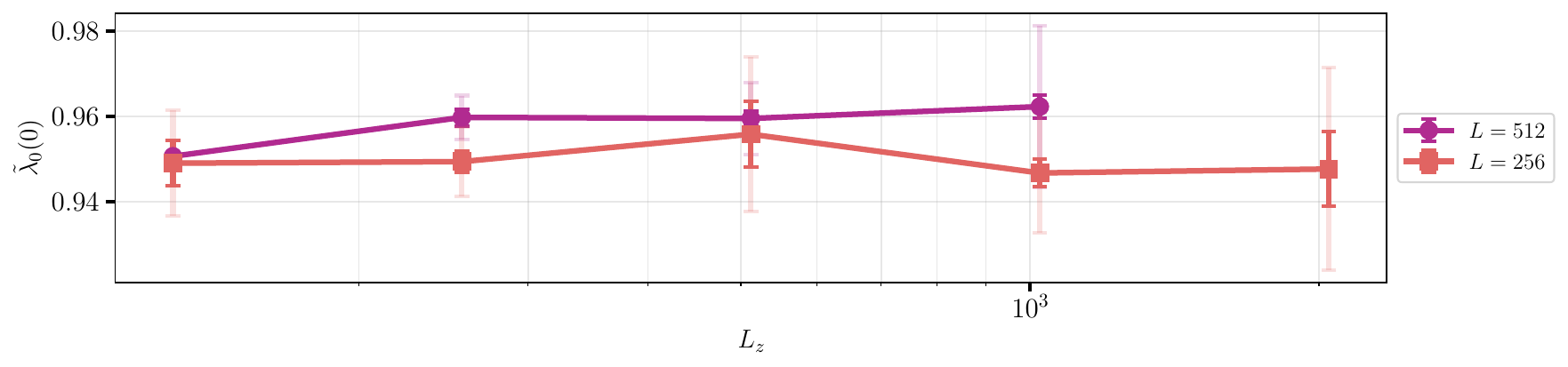}
  \caption{Volume dependence of $\tilde\lambda_0(0)$ (with $V=L^2L_z$), obtained by averaging $\lambda(t)$ over the interval defined in the text. Dark error bars: statistical uncertainty of the average. Light error bars: max-min spread of $\tilde \lambda_0(t)$ over the averaging window. $L=48$ is excluded because the interval is empty.
  \vspace{2em}}

    \label{fig:lambda}
\end{figure}

As a representative odd operator we use the spin field (lattice spins $s_i=\pm1$) and for even operators, the local energy density
\begin{equation}
\epsilon(j)=\frac{1}{6}\sum_{(i,j)} J_{ij}\,s_j s_i\,,
\end{equation}
with $(i,j)$ the six nearest neighbors of $j$. Throughout this section $m$ denotes the lightest mass, which is the inverse of the correlation length. Correlation functions are measured using FFT. 

\subsection{1pt functions}

As discussed in sec.~\ref{sec:correlation functions}, the wall-induced one-point shift is not universal: it depends on the chosen lattice discretization through the delta residue and through UV-sensitive contributions in the spectral density $\tilde\rho$. We normalize the delta by extracting the overlap $\mathcal{N}_\epsilon$ from the long-distance behavior of the periodic connected two-point function,
\begin{equation}
    \langle \epsilon(r)\epsilon(0)\rangle_\textrm{P}^\text{conn} \simeq \frac{\mathcal N_\epsilon^2}{4\pi}\,\frac{e^{-mr}}{r}\,.
\end{equation}
The critical scaling of $\mathcal N_\epsilon$ can be determined as follows.
Consider the lattice correlator at distance $r=1/m$,
\begin{equation}
   \langle 
\epsilon(r=1/m)\epsilon(0)\rangle_\textrm{P}^\text{conn}
\sim  m \,\mathcal N_\epsilon^2 
\end{equation}
Now let us compare this with the UV CFT prediction,
\begin{equation}
\langle \epsilon(r)\epsilon(0)\rangle_\textrm{P}^\text{conn} \sim  a^{2\Delta_\epsilon} 
\langle \epsilon_{\rm CFT}(r)\epsilon_{\rm CFT}(0)\rangle
 = \frac{a^{2\Delta_\epsilon}}{r^{2\Delta_\epsilon}}\,,
\end{equation}
where $a$ denotes the lattice spacing.
Comparing the two predictions for $r=1/m$, we find\footnote{Recall the basic relations $ma\sim t^\nu$ with $t=(T_c-T)/T_c$ and $\Delta_\epsilon=3-1/\nu$.}
\begin{equation}
\mathcal N_\epsilon^2 \sim a(ma)^{2\Delta_\epsilon-1} \sim a \,t^{\nu(2\Delta_\epsilon-1)}\sim a\, t^{1.1498549(4)}\,,
\end{equation}
using the values quoted in \cite{Chang:2024whx}.
The numerical results are summarized in app.~\ref{app:amplitude}; we find $\mathcal{N}_\epsilon(t)=5.7(4)\sqrt{a}\,t^{0.56(1)}$.

A similar scaling estimate applies to the wall-core strength: the wall profile $p_\epsilon(x_\perp)$ is controlled by the same scaling dimension $\Delta_\epsilon$ and, near criticality, takes the form $p_\epsilon(x_\perp)\sim (a/\xi)^{\Delta_\epsilon}h(x_\perp/\xi)$ with a universal scaling function $h$. Integrating over $x_\perp$ therefore gives $\mathcal{T}_{\rm lat}\equiv\int dx_\perp\,p_\epsilon(x_\perp)\sim a(\xi/a)^{1-\Delta_\epsilon}$, i.e.
\begin{equation}
\mathcal{T}_{\rm lat}(t)\sim a\,t^{\nu(\Delta_\epsilon-1)}
=a\,t^{0.26\dots}
\,,\label{eq:scalling t lat}
\end{equation}
again up to a non-universal prefactor and scaling corrections. 

We then define the dimensionless combination
\begin{equation}\label{eq:conversionspectral}
    \tilde \lambda_0 \equiv %
    \dfrac{\sqrt m\mathcal{T_\epsilon^\textrm{lat}}}{ \mathcal{N}_\epsilon}
    \sim t^0\,,
\end{equation}
that remains finite when $t\to 0$.

We measure $\delta\langle \epsilon\rangle$ defined in eqs.~\eqref{eq:eps_1pt_ToverLz} and \eqref{eq:delta epsilon EFT}, and convert it to $\tilde\lambda_0$ using eq.~\eqref{eq:conversionspectral}. Figure~\ref{fig:1ptfunction z2 even} plots $\tilde\lambda_0$ as a function of $\min(L,L_z)/\xi(t)$ for several transverse sizes $L_z$ and wall areas $A=L^2$, so that the onset of finite-size effects is expressed directly in units of the bulk correlation length.

The data show that the leading $L_z$ dependence of the one-point function is the kinematic factor $1/L_z$ predicted by eq.~\eqref{eq:eps_1pt_ToverLz}. After this factor is removed through the definition of $\tilde\lambda_0$, the remaining variation with $L_z$ is comparable to the error bars. In other words, the dominant transverse-size dependence is accounted for by zero-mode delocalization.

As $\xi$ becomes comparable to $L$, $\tilde\lambda_0$ decreases and eventually becomes consistent with zero. This is expected: once $\sigma L^2\sim 1$, the domain wall melts and the distinction between the anti-periodic and periodic sectors is washed out. Accessing the scaling regime therefore requires both $\sigma L^2\gg 1$ and $mL_z\gg 1$.

In \cite{Lima:2025sqa} we found that higher-derivative EST corrections are small once $\sigma A\gtrsim 30$. As a reference, at $t=7\times10^{-3}$ (where $\xi\simeq 6$) one finds $\sigma A\simeq 4,\,123,\,492$ for $L=48,\,256,\,512$, respectively; we therefore exclude $L=48$ from the analysis. For $L=256$ and $L=512$, the lower end of the fitting window is set by $\sigma A\simeq 30$, corresponding to $t\simeq 1.6\times10^{-3}$ and $t\simeq 5.6\times10^{-4}$.

Fitting the largest volumes, $L=512$ and $L_z\in[512,1024]$, to $\tilde\lambda_0(t)=C\,t^\alpha$ over the interval $t\in[t_{\sigma A=30},\,t_{\xi=6}]$, with $t_{\xi=6}=7\times10^{-3}$ and $L_z\gtrsim 3\xi$, we find $\alpha$ statistically consistent with zero. The analogous fit for the integrated wall-core amplitude gives
\begin{equation}
\mathcal{T}^{\rm lat}_\epsilon(t)=2.6(2)\,t^{0.25(2)},
\end{equation}
in agreement with the scaling expected from eq.~\eqref{eq:scalling t lat}.

Finally, fig.~\ref{fig:lambda} shows $\tilde\lambda_0$ averaged over this temperature window for each geometry. The dominant residual dependence is on the wall area, while the remaining $L_z$ dependence is comparable to the error bars.

\subsection{2pt functions}
\label{subsec:two-point-functions}
We focus on $\beta=0.222136$ ($\xi \sim 12$) and $V=384^2\times256$. The qualitative behavior is unchanged across our datasets provided $L,L_z\gg \xi \gg a$. We restrict to separations along the wall in a single direction, denoted $x_\parallel$.

\subsubsection{\texorpdfstring{$\mathbb{Z}_2$}{Z2}-odd operators}
\label{subsubsec:z2-odd fields}
\begin{figure}
    \centering
    \includegraphics[width=\linewidth]{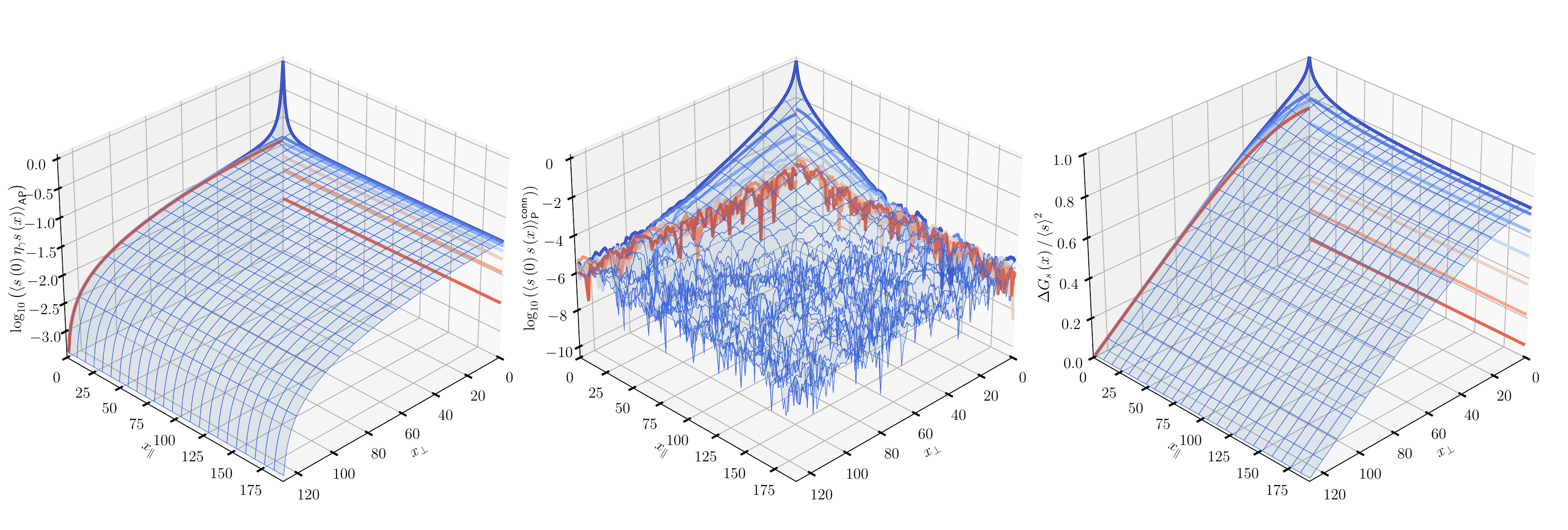}
\caption{Spin-sector correlators at $\beta=0.222136$ on $V=384^2\times256$, shown as functions of $x=(x_\parallel,x_\perp)$ in log10 scale. Left: $\langle s(0)\,\eta_\gamma\,s(x)\rangle_{AP}$ defined in \eqref{eq:s-eta-s}. Middle: $\langle s(0)s(x)\rangle^{\rm conn}_{P}$. Right: $\Delta G_s(x)$. Vertical planes show fixed-$x_\perp$ (or fixed-$x_\parallel$) slices of the surface.}

    \label{fig:spin 2pt functions}
\end{figure}
Eq.~\eqref{eq:2pt-odd-univ} predicts the shape of the normalized correlator difference $\Delta G_s(x)/\langle s\rangle^2$. For fixed $x_\parallel$ and $|x_\perp|\gg d(x_\parallel)$, it becomes approximately linear in $|x_\perp|$, and it vanishes at $|x_\perp|=L_z/2$. This behavior is visible in the right panel of fig.~\ref{fig:spin 2pt functions}.

\begin{figure}
            \centering
\includegraphics[width=0.495\linewidth]{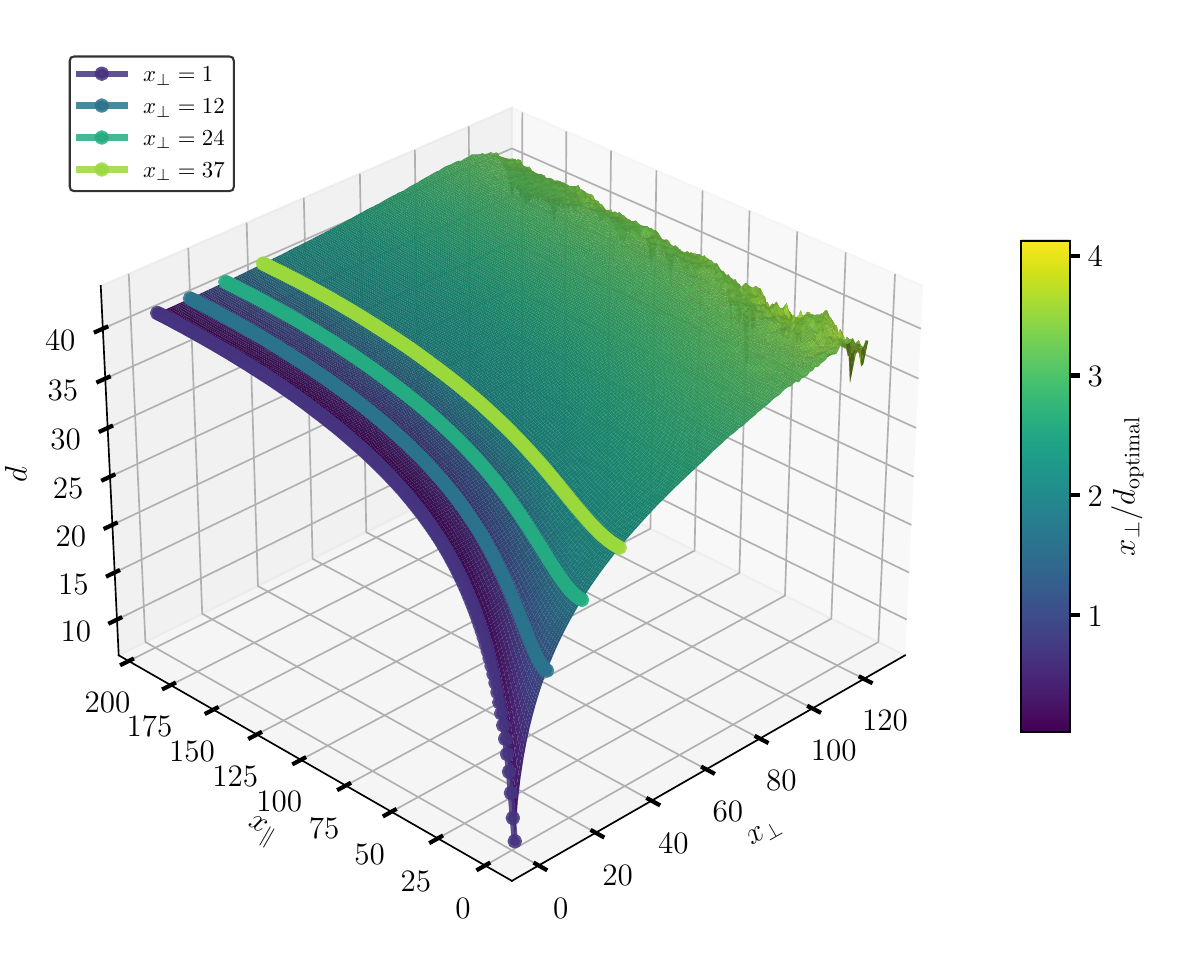} \includegraphics[width=0.495\linewidth]{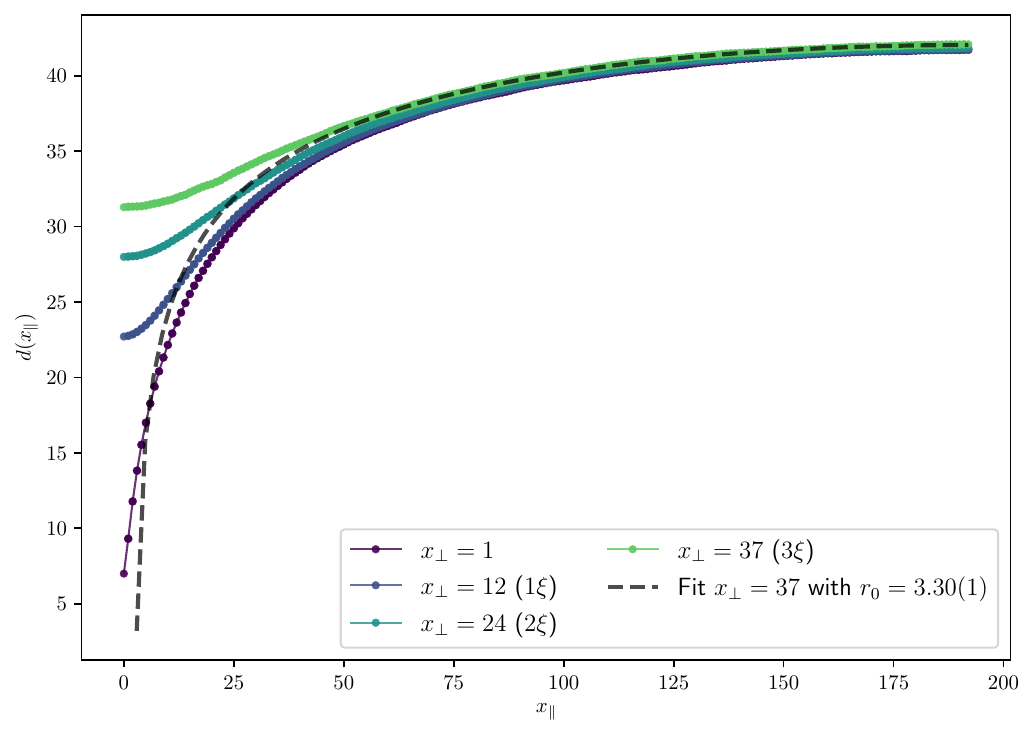}
   \caption{Extraction of the wall variance from the odd-sector correlator. Left: $d(x_\parallel,x_\perp)$ obtained by solving eq.~\eqref{eq:d_rootfinding} pointwise in $(x_\parallel,x_\perp)$. The highlighted curves are the fixed-$x_\perp$
 slices shown in the right panel. Right: $d^2(x_\parallel)$ at selected transverse positions $x_\perp$ compared to the free massless-boson prediction on a torus, eq.~\eqref{eq:d2_torus}; the fit determines the UV scale $r_0$ in eq.~\eqref{eq:r0}.}

    \label{fig:branon propagator}
\end{figure}

To extract $d(x_\parallel,x_\perp)$, we solve
\begin{equation}
\dfrac{\Delta G_{s}(x)}{\langle s \rangle^2}=1-\frac{2\,d(x_\parallel,x_\perp)}{L_z}\,
f\!\left(\frac{|x_\perp|}{d(x_\parallel,x_\perp)}\right)\,,
\label{eq:d_rootfinding}
\end{equation}
for $d(x_\parallel,x_\perp)$ for each value of $(x_\parallel,x_\perp)$. The results are shown in fig.~\ref{fig:branon propagator}. In the regime $L_z\gg |x_\perp|$ and $L_z\gg d(L/2)$, the extracted $d(x_\parallel,x_\perp)$ is only mildly dependent on $x_\perp$. 

We compare the measured variance $d$ to the free massless-boson prediction on a finite torus. 
We define $d^2(x_\parallel)\equiv\langle (\pi(x_\parallel)-\pi(0))^2\rangle$ by expanding the square and using point  splitting,
\begin{equation}
d^2_L(x_\parallel)
\equiv
2\lim_{y\to 0}
\left[
\langle \pi(y)\pi(0)\rangle
-
\langle \pi(x_\parallel)\pi(0)\rangle
+\frac{\alpha'}{2}\ln\frac{y^2}{r_0^2}
\right] .
\label{eq:d2_pointsplit}
\end{equation}
The logarithm subtracts the universal short-distance singularity of the free boson
propagator, while the non-universal length scale $r_0$ parametrizes the remaining finite
additive constant. 

Using the free boson Green function on a torus \cite{Polchinski:1998rq}, one obtains
\begin{equation}
d_L^2(x_\parallel)
=
2\alpha'
\ln\!\left[
\frac{L}{\pi r_0}
\left|
\frac{\vartheta_1\!\left(\frac{\pi x_\parallel}{L}\middle|q\right)}
{\vartheta_1'(0|q)}
\right|
\right],
\qquad q=e^{-\pi},
\qquad
\alpha'=\frac{1}{2\pi\sigma}.
\label{eq:d2_torus}
\end{equation}
For fixed $x_\parallel$ and $L\to\infty$, this reduces to
\begin{equation}
d^2_L(x_\parallel)=2\alpha'\ln\frac{|x_\parallel|}{r_0}
+O\!\left(\frac{x_\parallel^2}{L^2}\right).
\label{eq:d2_inf}
\end{equation}

\begin{figure}[t]
    \centering
    \includegraphics[width=0.44\linewidth]{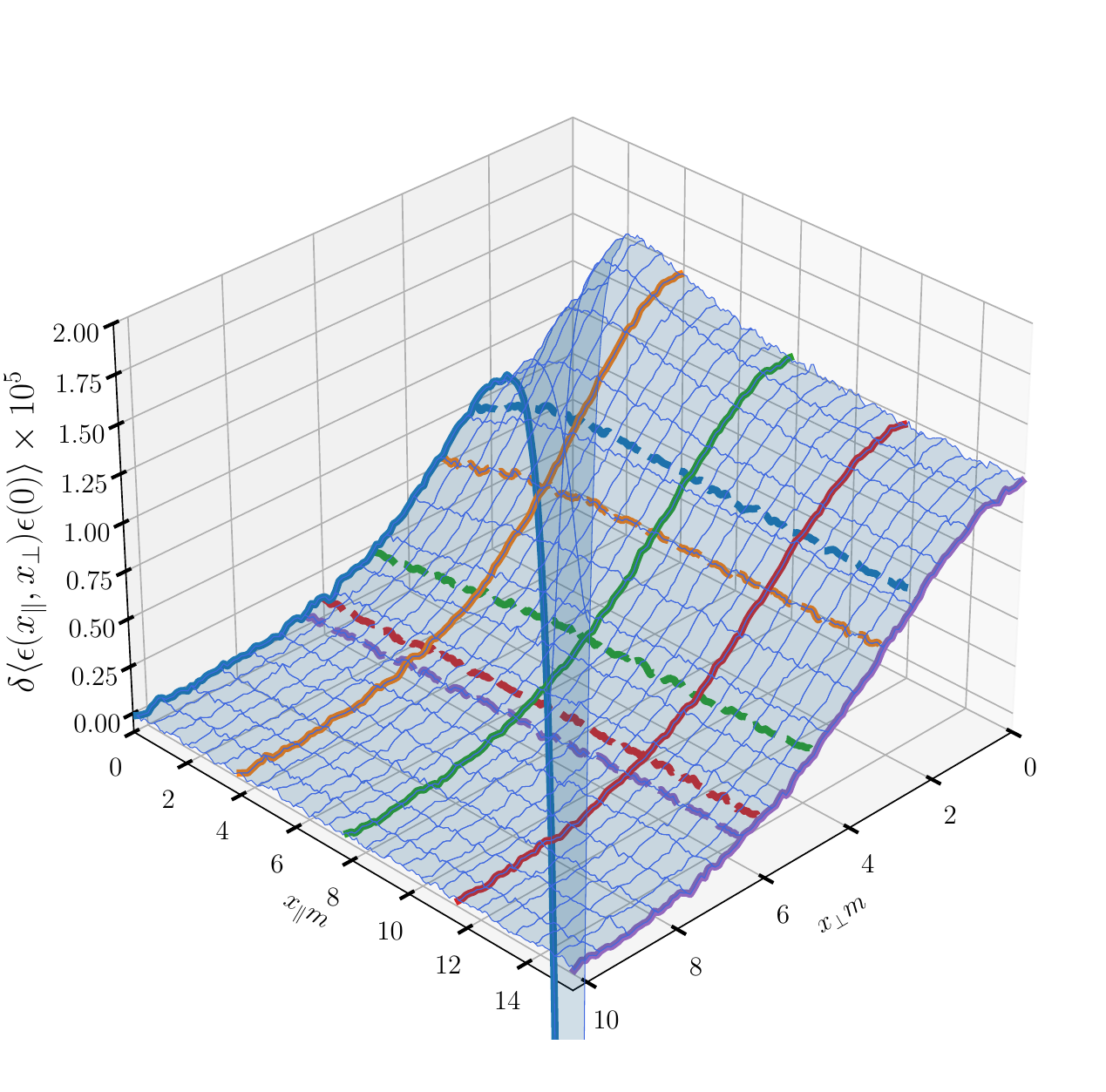}\includegraphics[width=0.44\linewidth]{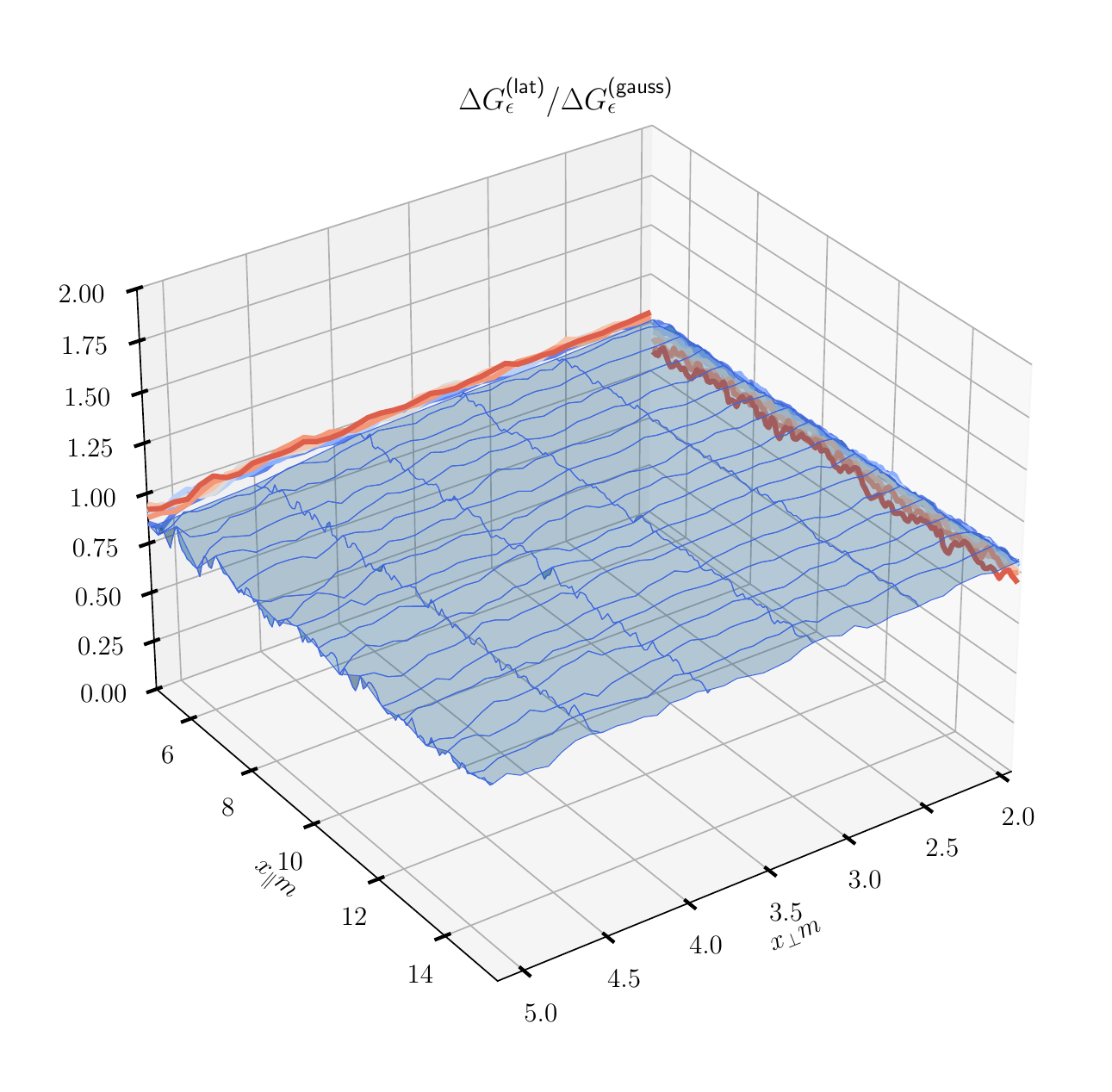}
    \caption{Even-sector energy correlator difference at $\beta=0.222136$ on $V=384^2\times256$. Left: $\Delta G_\epsilon(x)\equiv
        \langle\epsilon(0)\epsilon(x)\rangle^{\rm conn}_{AP}-\langle\epsilon(0)\epsilon(x)\rangle^{\rm conn}_{P}$. Solid lines denote the fixed-$x_\parallel$ cuts shown in left panel of fig.~\ref{fig:2ptlongdistanceregime}, while dashed lines denote the fixed-$x_\perp$ cuts shown in the left panel of fig.~\ref{fig:2ptlongdistanceregimexparallel}. 
        Right: Comparison between MC data and the Gaussian prediction in eq.~\eqref{eq:2ptover1ptsquared} and eq.~\eqref{eq:DeltaGeps_gaussian}  for $x_\perp\in[2\xi,L_z/4]$ and $x_\parallel\gtrsim 5\xi$.}
    	    \label{fig:energy 2pt functions correction}
        \vspace{0.2cm}
	    \includegraphics[width=0.88\linewidth]{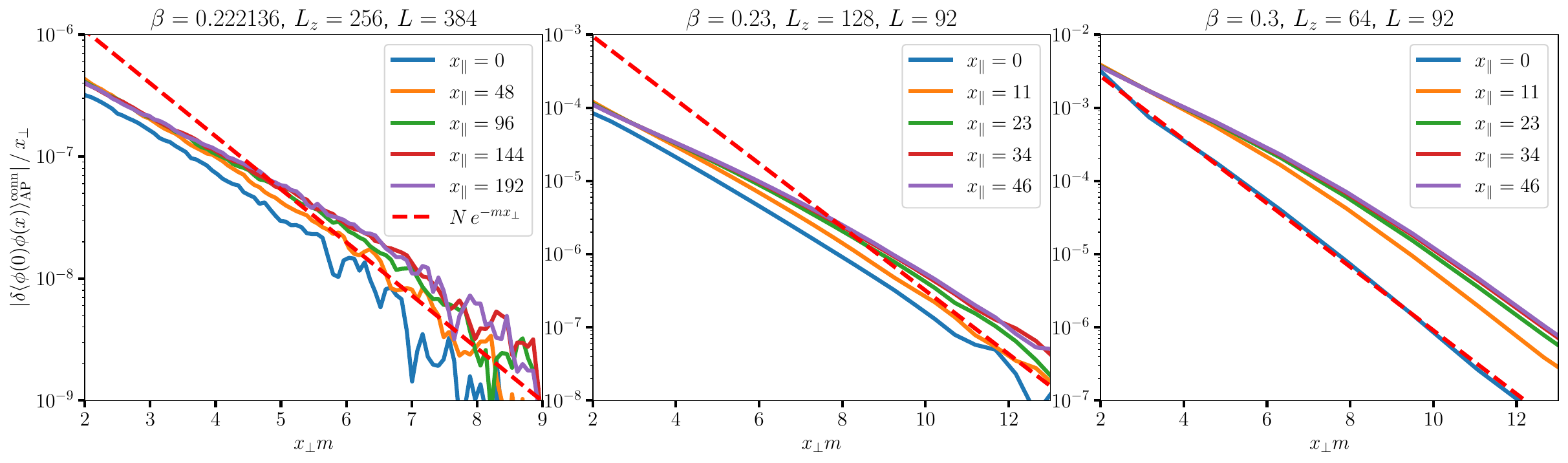}
	    \caption{ Long-distance regime of the even-sector correlator difference. We plot $|\delta\langle\epsilon(0)\epsilon(x_\parallel,x_\perp)\rangle|/|x_\perp|$ as a function of $m x_\perp$ for several $x_\parallel$ (legend) at the indicated $\beta$ and volumes. Dashed lines show a reference exponential $\propto e^{-m x_\perp}$.}
	    \label{fig:2ptlongdistanceregime}
        \vspace{0.2cm}
	
	    \includegraphics[width=0.88\linewidth]{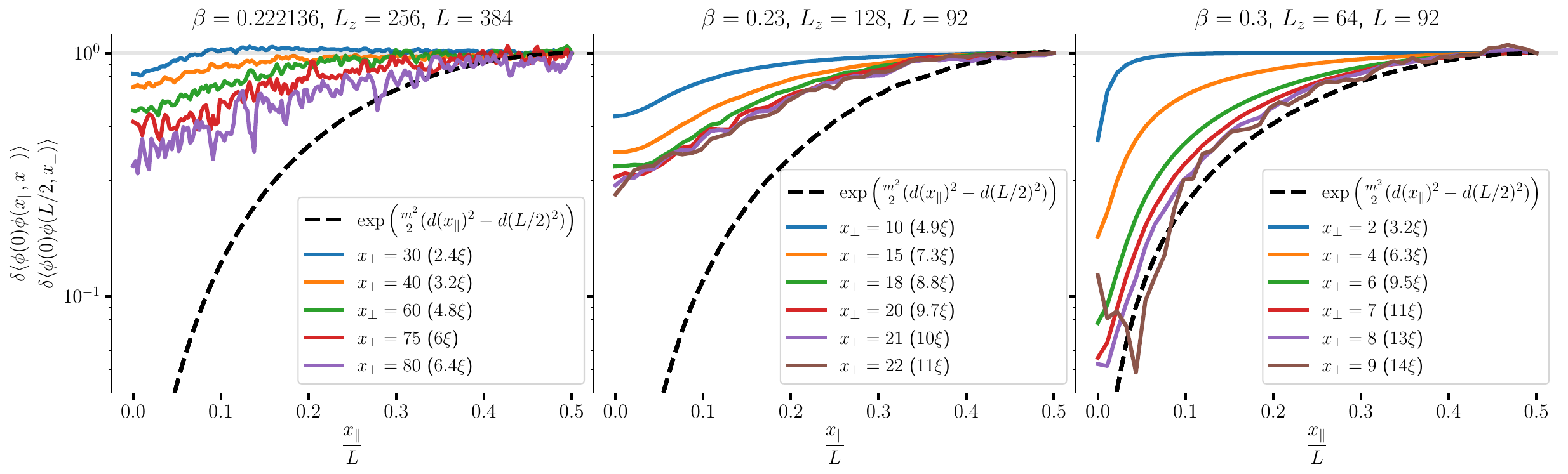}
	    \caption{Test of the predicted $x_\parallel$ dependence in the long-distance regime. We plot $\delta\langle\epsilon(0)\epsilon(x_\parallel,x_\perp)\rangle/\delta\langle\epsilon(0)\epsilon(L/2,x_\perp)\rangle$ as a function of $x_\parallel/L$ for several $x_\perp$ (legend), at the indicated $\beta$ and volumes. Dashed lines show the prediction for the $d^2(x_\parallel)$ extracted from the spin 2pt functions.
    }
\label{fig:2ptlongdistanceregimexparallel}
	\end{figure}

The dashed line in fig.~\ref{fig:branon propagator} is obtained by fitting $r_0$ using $x_\parallel>150$; the fit remains consistent down to $|x_\parallel|\sim 50\approx 4\xi$. For selected $x_\perp$ we obtain
\begin{equation}
r_0\big|_{x_\perp=1}^{V = 384^2 \times 256} =3.50(1)a,\qquad
r_0\big|_{x_\perp=\xi}^{V = 384^2 \times 256}=3.49(1)a,\qquad
r_0\big|_{x_\perp=2\xi}^{V = 384^2 \times 256}=3.43(1)a.
\label{eq:r0}
\end{equation}

\subsubsection{\texorpdfstring{$\mathbb{Z}_2$}{Z2}-even operators}
\label{subsubsec:z2-even fields}
The left panel of fig.~\ref{fig:energy 2pt functions correction} shows
$\Delta G_\epsilon(x_\parallel,x_\perp)$, i.e.\ the even-sector correlator
difference (equivalently $\delta\langle\epsilon(0)\epsilon(x)\rangle$). The Gaussian
profile predicted by the nearby-regime expression \eqref{eq:2ptover1ptsquared}
is visible in the fixed-$x_\parallel$ transverse slices shown in that panel:
as a function of $x_\perp$, the correction is approximately Gaussian. To test
this quantitatively, the right panel shows the ratio of the Monte Carlo data to
the prediction of eq.~\eqref{eq:2ptover1ptsquared}, so that agreement with the
Gaussian form appears as a plateau near $1$. The ratio remains close to $1$ for
$x_\perp\lesssim 5\xi\sim 60$ over the measured range of $x_\parallel$.

Fig.~\ref{fig:2ptlongdistanceregime} compares $\delta\langle\epsilon(0)\epsilon(x)\rangle/|x_\perp|$ to the  exponential falloff at large $x_\perp$ predicted by eq. \eqref{eq:2ptlambda0exponential}. The overall coefficient is not extracted because the predicted $x_\parallel$ dependence cannot be cleanly resolved, as shown in fig.~\ref{fig:2ptlongdistanceregimexparallel}, where we represent
\begin{equation}
    \dfrac{\delta \langle \epsilon(0)\epsilon(x_\parallel,x_\perp)}{\delta \langle \epsilon(0)\epsilon(L/2,x_\perp)}\approx e^{-\frac{1}{2}m^2\left(d^2(L/2) - d^2(x_\parallel)\right)},
\end{equation}
for several values of $x_\perp$. The decrease with $x_\perp$ is suggestive, but larger $x_\perp$ is required for a quantitative test. The dataset at $\beta=0.3$ is closer to the predicted behavior; however, despite being below the roughening transition, $\beta_c^\text{R} = 0.40754(4)$ \cite{Hasenbusch:1996fd,PhysRevB.42.545}, discretization effects are not under control there.

\subsection{Free energies}
\label{sec:freeenergies}

We finally consider the finite-$L_z$ dependence of the effective string tension. 
The free-energy difference is determined as follows. We simulate the 3D Ising model in an extended ensemble where the spin configuration $\{s\}$ and the plane coupling $J$ are sampled jointly. A fixed plane $\Sigma$ is chosen, and all bonds crossing $\Sigma$ are multiplied by $J$, with $J=1$ and $J=-1$ corresponding to the periodic and anti-periodic sectors. For fixed $J$,
\[
e^{-F(J)} \equiv \sum_{\{s\}} e^{-\beta H[\{s\};J]},
\]
so that $F(1)=-\log Z_{\rm P}$ and $F(-1)=-\log Z_{\rm AP}$.

To sample efficiently in $J$, we introduce a bias potential $\omega(J)$ and simulate
\[
\mathcal Z_{\rm ext}(\beta)=\int_{J_{\min}}^{J_{\max}} dJ\, e^{-F(J)+\omega(J)}.
\]
The bias is constructed by a flat-histogram Wang--Landau procedure \cite{Berg:1991cf,Lee:1993cca,Wang:2000fzi,Trebst_2004,Viana_Lopes_2006}, following the implementation of \cite{Lima:2025sqa}. After convergence, $\omega(J)=F(J)+\text{const}$, and therefore
\[
\Delta F_{\mathbb Z_2}=F(-1)-F(1)=\omega(-1)-\omega(1).
\]
This gives direct access to the twisted-sector free-energy difference, without reconstructing it from $L_z$ derivatives or separate bulk-subtracted free energies.

A quantitative test of the asymptotic prediction is beyond the precision currently available. There are two separate reasons. First, the signal itself is exponentially small in the regime $mL_z \gg 1$ in which the asymptotic formula is derived. Second, even within the EFT, the approach to that regime is slow, as encoded in the pre-asymptotic factor $F(mL_z)$ introduced in eq.~\eqref{eq:sigmaeffcorrected}. The purpose of the present analysis is therefore limited: we ask whether the sign, scale, and qualitative $L_z$ dependence of the Monte Carlo data are consistent with the predicted large-$L_z$ behavior, but we do not attempt to extract $\lambda$ from these data.

\begin{figure}
    \centering
    \includegraphics[width=\linewidth]{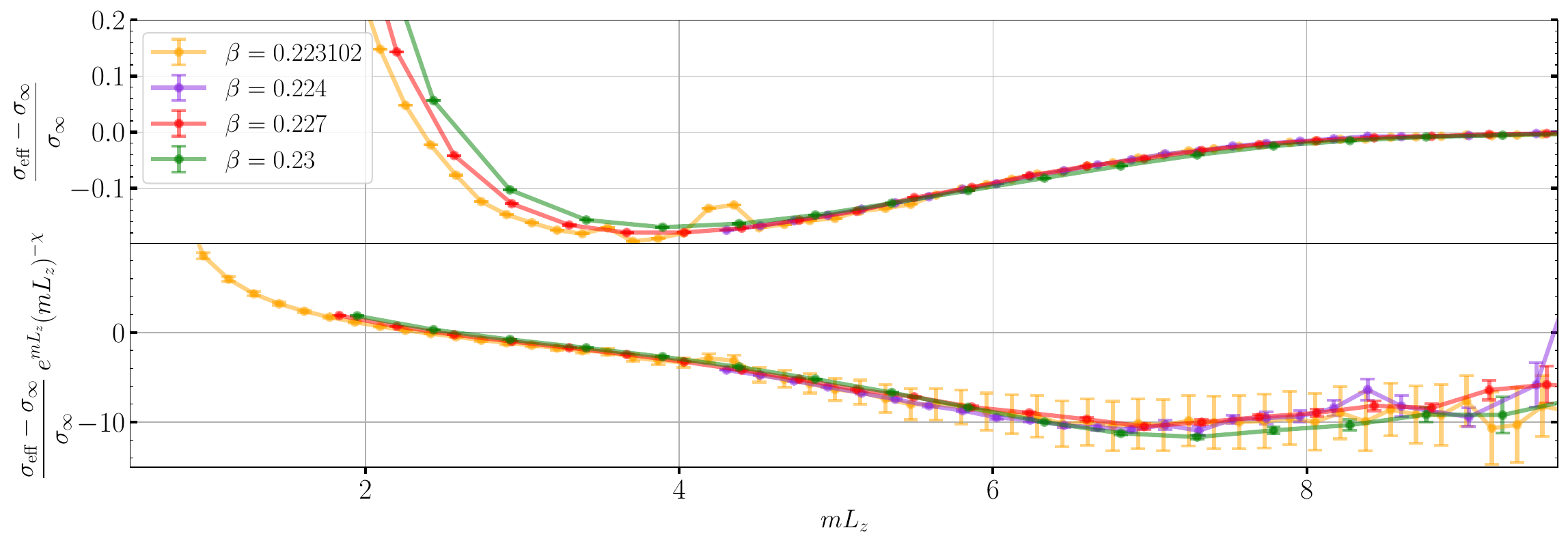}
    \caption{Finite-$L_z$ shift of the effective string tension. Top: Monte Carlo data for $(\sigma_{\rm eff}(L_z)-\sigma_\infty)/\sigma_\infty$ at the indicated values of $\beta$. Bottom: the same data rescaled by $e^{mL_z}(mL_z)^{-\chi}$, with $\chi=m^2/(4\pi\sigma)$. For $mL_z\gtrsim 3$, the shift is negative and approximately follows the predicted leading dependence $-(mL_z)^\chi e^{-mL_z}$, while the residual drift in the lower panel shows that sizable pre-asymptotic corrections remain.}
    \label{fig:free energy finte volume corrections}
    \centering
\end{figure}

Figure~\ref{fig:free energy finte volume corrections} shows the Monte Carlo determination of
\begin{equation}
\frac{\sigma_{\rm eff}(L_z)-\sigma_\infty}{\sigma_\infty}
\end{equation}
for four values of $\beta$, in the range where the extrapolation to infinite wall area is under control. Here $\sigma_\infty$ denotes the large-$L_z$ limit of the effective string tension at fixed temperature. For $mL_z \gtrsim 3$, the shift is negative, as expected from the prediction of sec.~\ref{sec: finite transverse volume corrections}. Moreover, when plotted as a function of the scaling variable $mL_z$, the data for different $\beta$ approximately collapse onto a common curve. This is consistent with the expectation that the leading dependence is governed by the universal
combination $(mL_z)^\chi e^{-mL_z}$, with
\begin{equation}
\chi=\frac{m^2}{4\pi\sigma}\, .
\end{equation}

The lower panel of fig.~\ref{fig:free energy finte volume corrections} shows the same data rescaled by the leading asymptotic kinematic factor,
\begin{equation}
e^{mL_z}(mL_z)^{-\chi}\,
\frac{\sigma_{\rm eff}(L_z)-\sigma_\infty}{\sigma_\infty}\, .
\end{equation}
If the asymptotic regime were already reached, this quantity would approach a constant proportional to $-\lambda^2$. Instead, the residual drift remains significant throughout the accessible range. This is expected. As already indicated by the numerical evaluation of the pre-asymptotic function $F(l)$ in sec.~\ref{sec: finite transverse volume corrections}, the convergence to the asymptotic form is slow: for the representative value of $r_0$ extracted from the two-point analysis, one finds $F(mL_z)\simeq 0.7$ even at $mL_z\simeq 30$. Sizeable pre-asymptotic corrections are therefore still present in the region accessible to the Monte Carlo simulation. At present, the free-energy observable is best viewed as supporting the sign and leading kinematic structure of the EFT prediction, but not as an independent precision probe of the coupling.

\section{Discussion}
\label{sec:discussion}

 We formulate an effective description of the coupling between a fluctuating domain wall and the lightest bulk massive mode, in the regime where the relevant observables are dominated by long-wavelength dynamics along the wall, equivalently by nearly on-shell bulk exchange carrying small parallel momentum. 

Observables are computed perturbatively in the wall-particle coupling, but must remain non-perturbative in the wall fluctuations themselves: a naive rigid-wall expansion misses the logarithmic growth of the branon variance, whose resummation generates a non-trivial kinematic enhancement governed by the exponent $\chi = m^2/(4\pi\sigma)$. As a result, the leading finite-$L_z$ correction to the wall free energy scales as $(mL_z)^\chi e^{-mL_z}$, while the asymptotic even-sector two-point function behaves as $(m|x_\parallel|)^{2\chi} e^{-m|x_\perp|}$, rather than with the conventional rigid-wall form $e^{-mx_\perp}$. 

The amplitude of these effects is controlled by a renormalized dimensionless coupling $\lambda$, rather than directly by the bare coupling in the action; the same parameter also governs the low-energy analytic structure of particle scattering off the wall. 

Numerically, the Monte Carlo data supports the nearby-regime predictions, including the kinematic $1/L_z$ scaling of one-point functions, the extraction of the wall variance from odd-sector correlators, and the Gaussian transverse profile of even-sector correlators, while the enhanced asymptotic regime for the two-point functions is only seen qualitatively at present, thus the precision is not yet sufficient for a  determination of the coupling $\lambda$. A quantitative test of the free-energy correction would require values of $mL_z$ beyond what is likely to be accessible in the near future.

This EFT can alternatively be thought of as a non-relativistic EFT (see for instance \cite{Guth:2014hsa}). Suppose that we perform a Wick rotation so that time is the direction orthogonal to the domain wall. Then, we can write the relativistic massive field in the bulk as
$\phi= \frac{1}{\sqrt{2m}}\left(e^{-im t} \Psi + e^{im t} \Psi^*\right)$.\footnote{We thank G. Cuomo for suggesting this point of view.}  In this frame, the interaction term \eqref{eq:interaction} becomes
\begin{equation}
S_{\textrm{int}}\sim \frac{\lambda_0}{\sqrt{2m}} \int d^{2}x \left[ e^{-im \pi(x)} \Psi + e^{im \pi(x)} \Psi^* \right]\,,
\end{equation}
where $\Psi$ is a slow-varying non-relativistic field associated with the bulk particle. From this perspective, the bulk particle interacts with the Goldstone bosons $\pi$ at a fixed moment in time and through the vertex operators $e^{\pm im \pi(x)}$, with scaling dimension $\chi=\frac{m^2}{4\pi \sigma}$.
It would be interesting to develop this further to systematize the power counting of this EFT.

Our EFT approach to the coupling between bulk massive particles and a fluctuating two-dimensional surface can also be applied to the interactions between glueballs and flux tubes in confining gauge theories. For example, one can compute the one-point function of a local bulk operator in the presence of two Polyakov loops in a confining gauge theory. This setup has recently been studied in \cite{Caselle:2026coc,Verzichelli:2026ipr}. 
More precisely, one considers two straight Polyakov loops at distance $R$ from each other. In the confining phase, there is a flux tube whose worldsheet is bounded by the Polyakov loops. Then, we can measure the one-point function of a local bulk operator placed at the same distance from both Polyakov loops. Its position is characterized by the distance $y$ to the plane containing the Polyakov loops. 
Let us denote by $\langle \phi(0,y) \rangle_{\rm Pol}$ the one-point function in the presence of the Polyakov loops and by $\langle \phi \rangle$ the one-point function without other insertions.
For large $y$, we can use our effective field theory as follows
\begin{align}
    \langle \phi(0,y) \rangle_{\rm Pol}-\langle \phi \rangle &\approx -\lambda_0 \int d^2x\langle \phi(x,\pi(x)) \phi(0,y) \rangle \\
    &\approx -\lambda_0 \int d^2x\left\langle \frac{e^{-m\sqrt{x^2+(y-\pi(x))^2}}}{4\pi\sqrt{x^2+(y-\pi(x))^2}} \right\rangle\\
   &\approx -\frac{\lambda_0}{2m} e^{-m|y|}  \left\langle  e^{-m\pi(0) } \right\rangle =
   -\frac{\lambda_0}{2m} e^{-m|y|}    e^{\frac{1}{2}m^2\langle \pi^2(0)\rangle } \\
  & = {\rm const}\, R^{\chi}\, e^{-m|y|} \,,
\end{align}
where we used  $\langle \pi^2(0)\rangle \approx \frac{1}{2\pi\sigma} \log R/R_0$ for
the central variance of the flux tube \cite{Luscher:1980iy, Caselle:1995fh}. $R_0$ is a UV scale different from $r_0$ in \eqref{eq:variance} because it also receives contributions from the boundary (Polyakov loops).
It would be interesting to check this prediction against the Monte Carlo simulations of $SU(2)$ lattice gauge theory in \cite{Caselle:2026coc,Verzichelli:2026ipr}.

\acknowledgments
We thank M. Caselle,  G. Cuomo, V. Gorbenko, A. Guerrieri, A. Hebbar and M. Meineri for useful discussions. JM is grateful for the feedback from participants of the conference "Bridging analytical and numerical methods for quantum field theory". 
This work was supported by FCT - Fundação para a Ciência e Tecnologia, I.P. by project reference and DOI identifier \href{https://doi.org/10.54499/2021.04743.BD}{https://doi.org/10.54499/2021.04743.BD}. 
JM and JV thank the cluster time provided by INCD funded by FCT and FEDER under the grants 2021.09830.CPCA, 2023.11029.CPCA, 2024.09383.CPCA  as well as GRID FEUP. They also thank Centro de Física do Porto funded by Portuguese Foundation for Science and Technology (FCT) within the Strategic Funding UIDB/04650/2020.
JP is supported by %
the Swiss National Science Foundation through the project 200020\_197160 and through
the National Centre of Competence in Research SwissMAP.

\appendix
\section{3D Ising}\label{sec:masses} \label{app:amplitude}

To determine the bulk inputs used in the main text, we analyze periodic connected two-point functions in the large-distance regime. For the energy operator, we fit
\begin{equation}
\langle \epsilon(r)\epsilon(0)\rangle_c \simeq \frac{\mathcal{N}_\epsilon^2}{4\pi}\,\frac{e^{-mr}}{r},
\label{eq:bulk_eps_fit}
\end{equation}
implemented by fitting $r\langle \epsilon(r)\epsilon(0)\rangle_c$ on a semilog plot. The lower fit bound $r_{\min}$ is chosen such that the best-fit curve, extrapolated to slightly smaller separations, still agrees with the data for a few points beyond the fit window, indicating that residual short-distance and excited-state effects are subleading at $r\ge r_{\min}$. The upper bound $r_{\max}$ is taken as the largest separation before the signal becomes noise-dominated, always with $r_{\max}\lesssim L/4$ to avoid wrap-around effects.

In the symmetry-broken phase, spin and energy correlators are governed by the same lightest mass gap. In practice, however, the spin-spin correlator has a cleaner overlap with the lightest state, so we extract $m$ from the spin channel and use eq.~\eqref{eq:bulk_eps_fit} only to determine $\mathcal{N}_\epsilon$. Since our goal is a robust input for the worldsheet analysis rather than a high-precision determination of bulk amplitudes, we use uncorrelated least-squares fits and quote statistical errors. As a basic robustness check, we repeat the fits with $r_{\min}\to r_{\min}\pm1$ and, when useful, with a nearby $r_{\max}$; the resulting variation is negligible or comparable to the statistical uncertainty.

From these bulk two-point functions we obtain $m$ and $\mathcal{N}_\epsilon$, while $\sigma$ is taken from ref.~\cite{Lima:2025sqa}. The results are collected in table~\ref{tab:my_label}. To convert $\mathcal{T}_\epsilon^{\rm lat}$ to $\lambda$ at the small reduced temperatures relevant for the one-point analysis, we need $m$ and $\mathcal{N}_\epsilon$ for several temperatures close to the critical point. To reduce the number of dedicated simulations, we fit their temperature dependence and use the resulting interpolation. The results, in lattice units, $a=1$, are
\begin{align}
    \mathcal{N}_\epsilon (t) &=5.7(4)t^{0.56(1)}\\
    \chi(t)&=0.79(1) -0.15(30)t^{\omega \nu}+ 1.7(1.0)t\\
    \sigma(t)&=1.515(2) t^{2\nu} \left(1  - 0.28(2)t^{\omega \nu} - 0.31(6)t\right)\\
    m(t)&=3.832(1)t^{\nu}.
\end{align}
where $\nu=0.62997097(12)$ \cite{Chang:2024whx}, $\omega=0.82951(61)$ \cite{Reehorst:2021hmp}.

\begin{table}[ht]
    \centering
    \footnotesize
    \begin{threeparttable}
    \begin{tabular}{|l|c|c|c|c|c|c|}
        \hline
        \diagbox[width=5em]{param}{$\beta$}& 0.222136 & 0.223102 & 0.224 & 0.227 & 0.2285 & 0.23 \\
        \hline
        $m$ &  0.083(4)& 0.161(1) & 0.2178(1) & 0.3665(5) & 0.4287(6) & 0.4867(2) \\\hline
        $\sigma$ \cite{Lima:2025sqa} &  ---& 0.0026062(12) & 0.0047567(10) & 0.0132004(25) & 0.0178901(14) & 0.0227944(20) \\\hline
        $\chi$ &  ---& 0.791(9) & 0.794(1) & 0.810(2) & 0.817(2) & 0.827(1) \\\hline
        $\mathcal{N}_\epsilon$ & 0.0027(3) & 0.0091(9) &  ---&  ---&  ---& 0.07(1) \\
        \hline
    \end{tabular}   
    \caption{Bulk input parameters used in the analysis. The mass gap $m$ is extracted from periodic spin-spin correlators, $\mathcal{N}_\epsilon$ from periodic energy-energy correlators, and $\sigma$ is taken from ref.~\cite{Lima:2025sqa}. A dash indicates that the corresponding quantity was not extracted at that value of $\beta$. The quantity $\mathcal{N}_\epsilon$ is determined only for the subset of $\beta$ values used in the quantitative matching tests and for a few additional points needed to fit its small-$t$ extrapolation for the one-point analysis.}\label{tab:my_label}

    \end{threeparttable}
\end{table}

\bibliographystyle{JHEP}
\bibliography{All.bib,ExtraBib}

\end{document}